\documentclass[a4paper, 12pt]{article}
\usepackage[utf8]{inputenc}
\usepackage{color}
\usepackage{hyperref} 
 \usepackage{latexsym}
  \usepackage{amssymb}
  \usepackage{graphicx}
   \DeclareGraphicsExtensions{.jpg .png .gif .pdf .pdf}
\usepackage[T1]{fontenc}
 \usepackage[english]{babel}
  \usepackage{}
  \usepackage{amstext}
  \usepackage{amsmath}
  \usepackage{times} 
  \usepackage{amsfonts}
 \date{}

\title{Regular and Chaotic Behaviors of  Modified Rayleigh-Duffing oscillator }
\author{C. H. Miwadinou$^1$\footnote{clement.miwadinou@imsp-uac.org, hodevewan@yahoo.fr}, 
A. V. Monwanou$^1$\footnote{movins2008@yahoo.fr}, C.Aïnamon$^2$\footnote{ainamoncyrille@yahoo.fr} and\\ 
J. B. Chabi Orou$^{1,2}$\footnote{Author to whom correspondence should be addressed: jchabi@yahoo.fr}}

\begin{document}

\maketitle 1 Institut de Math\'ematiques et de Sciences Physiques, BP: 613 Porto Novo, Bénin,

2 Ecole Doctorale Sciences des Matériaux, Université d'Abomey-Calavi, \\
Bénin.

\begin{abstract}
The regular and chaotic behavior of modified Rayleigh-Duffing oscillator is studied. We consider in this paper the dynamics of \\
Modified  Rayleigh–Duffing oscillator. The  harmonic balance method are used to find the amplitudes of the oscillatory states, and analyze.
 The influence of system parameters are clearly found on the bifurcations in the response of this system is investigated.
 It is found also hysteresis and jump phenomenon are appered or desappered when certain parameters
incrases or descrases. Various bifurcation structures, the variation of the Lyapunov 
exponent are obtained, using numerical simulations of the equations of motion. Various basin attraction are used to confirm the predictions of bifurcation
structures and its corresponds Lyapunov exponent.
\end{abstract}

\section{Introduction}

In recent years, a twofold interest has attracted theoretical, numerical, and experimental investigations to understand 
the behavior of nonlinear oscillators. Theoretical (fundamental) investigations reveal their rich and complex behavior,
 and the experimental (self-excited oscillators) describes the evolution of many biological, chemical, physical, mechanical,
 and industrial systems \cite{1,2,3}. 
The interest devoted to chaos by many scientist
is due to the fact that this new phenomenon appears in various fields, from mathematics,
physics, biology, and chemestry, to engineering, economics and medecine. Consequently,
there are many opportunities for  application of chaos. For example, in physics chaos 
has been used to refine the understanding of planetory orbits, to reconceptualize quantum
level processes, and to forecast the intensitynof solar activity. In engineering, chaos
has been used in bulding of better digital filters, and to model the structural dynamics in such
structures as buckling columns. In medecine, it has been used to study cardiac arrhythmias and patterns
of disease communication. In psychology, it has been used to study mood fluctuations, the operation of the
olfactory lobe during perception, and partterns of innovation in organizations. In economics
it is being used to find patterns and develop new types of econometric model for the stack market to variations
in coton prices. There are also many opportunities for exploitation of chaos: synchronized chaos,
mixing with chaos, encoding information with chaos, anti-control of chaos, tracking of chaos and 
targetting of chaos. 
An important class of systems in general and in particular oscillators who presented a complex or chaotic behavior 
can be determined on the basis of nonlinear damping. Such damping can, in some systems, change the
sign depending on velocity or displacement values, and provide excitation energy to the examined system. These, so
called, self-excited damping terms are often used to describe systems with dry friction, bearings lubricated by a thin
layer of oil, shimming in vehicle wheels or chatter in a cutting process \cite{1,2}, \cite{5}-\cite{8} and \cite{17,18}. 
In Ref. \cite{28}, the authors have studied with considerable detail the effects of the damping level
on the resonance response of the steady-state solutions and in the basin bifurcation patterns of the
escape oscillator. In particular they analyzed the effect of using different damping levels and how this
contributes to the erosion of the safe areas in phase space, and they also provide a comprehensive global 
picture of the main bifurcation boundaries.
More recently such a nonlinear
damping force has also been considered \cite{19} in modelling of a modern vehicle suspension system due to electro- or magneto-rheological
 fluid damping where it is causing a hysteretic effect. In this model \cite{19} the authors used a self-excited
term of the Rayleigh and the Duffing type with a double well potential. 
 Parametric excitation occurs in a wide variety of engineering application
 (Refs. \cite{10}-\cite{14}).
 In this vein, we propose to study in this paper regular and chaotic behavior of the modified Rayleigh-Duffing oscillator
 whose  equation is 
the form Eq. (\ref{eq.0}). This equation which having nonlinear dissipative terms and parametric excitation term can be used to model some systems
such as Brusselator, Selkov, rolling response, certain  $MEMS$ systems... (\cite{5}-\cite{8}, \cite{15}, and \cite{21}- \cite{24}).

The paper is organized as follows: In Section 2, we describe the model. Section 3 deals with the amplitude of the forced harmonic oscillatory states,
 using the harmonic balance method \cite{2} . Section 4 addresses the phase portraits, largest Lyapunov exponent and the bifurcation diagrams
 from which a concluding remark can be made in connection to the tendency of the system to have quasiperiodic, nonperiodic evolutions and chaotic
 one according to the choice of initial conditions which match with the basin of attraction found. The last section devoted to the conclusion will 
point out the contribution of nonlinear damping therms and parametric excitation term which modified Ordinary Rayleigh-Duffing oscillator in both regular
and chaotic behaviour of oscillations of this system.

\section{Model}
In this paper, we consider following  Modified Rayleigh-Duffing oscillator.
\begin{eqnarray}
 &&\ddot{x} + \epsilon\mu (1-\dot{x}^2)\dot{x}+\epsilon\beta\dot{x}^2+\epsilon k_1\dot{x}x+\epsilon k_2\dot{x}^2 x +(\gamma+\alpha\cos\Omega t)x\cr
&&+\lambda x^3=F\cos\Omega t, \label{eq.0}
\end{eqnarray}
where $\epsilon, \mu, \beta, k_1, k_2, \gamma, \lambda, F$ and $\Omega$ are parametrs. Physically, $\mu, k_2, \beta$ and $k_1$ represent respectively pure, unpure cubic and pure, unpure quadratic nonlinear damping
coefficient terms, $ \alpha$ and $F $ are the amplitudes of the parametric and external periodic forcing, and
$\sqrt{\gamma}$ and $\Omega$ are respectively natural and external forcing frequency. Moreover $ \lambda$ characterize the intensity of the nonlinearity and $\epsilon$ is the nonlinear damping parametr
control. The nonlinear damping term corresponds to the Modified Rayleigh
 oscillator, while the nonlinear restoring force corresponds to the Duffing oscillator. 

 This oscillator is used to modelize the following phenomena: 
A El Ni$\tilde{n}$o  Southern Oscillation $(ENSO)$ coupled tropical ocean-atmosphere weather phenomenon  in which the state
variables are temperature and depth of a region of the ocean called the thermocline (where the annual seasonal cycle is the parametric excitation and the 
model exhibits a Hopf bifurcation in the absence of parametric excitation) (\cite{21},  \cite{22}) ,  a $MEMS$ device  consisting 
of a $30\mu m$ diameter silicon disk which can be made to vibrate by heating it with a laser beam resulting in a Hopf bifurcation
 (where the parametric excitation is provided by making the laser beam intensity vary periodically in time) (\cite{23}, \cite{24}), the rolling response 
(\cite{5}-\cite{8}) etc. For examples, 
 the nonlinear ship rolling response and nonlinearly damped
universal uscape oscillator can be rewritten as follow:
\begin{eqnarray}
 &&\ddot{x} + \sum_{p=1}^nc_p\dot{x}|\dot{x}|^{p-1}+ \sum_{j=1}^ma_jx^j=F\cos\Omega t, \label{eq.00}
\end{eqnarray}
where $c_p$ is the nonlinear damping and $a_p$ restoring coefficients.\\
In the ship rolling case, A. Francescutto and G. Contento \cite{5} are used experimental results
and parameter identification technique to study bifurcations in ship rolling,  application of the extended Melnikov’s method are used by W. Wu and L
. McCue \cite{7} de  for single-degree-of-freedom vessel roll motion. In the other hand MIGUEL A. F. SANJU\`AN  (in Ref.\cite{29}) analyzed the effect of 
nonlinear damping on the universal escape oscillator. Another examples,
 consider the modified Rayleigh-Duffing oscillator equation which describes the glycolytic reaction catalized by phosphofructokinase, namely 
the Selkov equations  and  abstract trimolecular chemical reaction namely Brusselator oscillator \cite{15}:\\
One of these processes is simple classical two-variable model which describes
glycolytic reaction catalized by phosphofructokinase, namely the Selkov equations

\begin{equation}
 \left.
              \begin{array}{rl} 
            &  \frac{dx}{dt}=v-xy^2,\\
             & \frac{dy}{dt}=xy^2-wy,    
                 
                      \end{array}   
                         \right.     
   \end{equation}   
 and another is the Brusselator which describes an abstract trimolecular chemical
reaction
            \begin{equation}
 \left.
              \begin{array}{rl} 
            &  \frac{dx}{dt}=A+xy^2-(B+1)x,\\
             & \frac{dy}{dt}=Bx-xy^2,    
                 
                      \end{array}   
                         \right.     
   \end{equation} 
For the Selkov model $\lambda=v^2w^{-2}-w, \lambda''=(z_0w-3v)/w^2, \lambda'=w^{-2}, k=v^{-1}
\Omega=v/\sqrt{w}$, where $z_0=w^2/v+v/w.$ For the Brusselator $\lambda=1+A^2-B, \lambda''=
(B-2A^2)/A, \lambda'=1, k=A^{-1}, \Omega=A$ and $z_0=(B+A^2)/A.$
In both cases  $\xi=x+y-z_0$ is a deviation from the equilibrium concentration. With these 
conditions, the two last systems can been rewritten
\begin{equation}
 \frac{d^2\xi}{dt^2}+\tilde{\lambda}\frac{d\xi}{dt}+\tilde{\lambda}^{\prime\prime}(\frac{d\xi}{dt})^2+\tilde{\lambda}^{\prime}(\frac{d\xi}{dt})^3+
\Omega^2(1-k\frac{d\xi}{dt})^2\xi=0.
\end{equation}

 Perturbing this system by Duffing force, parametric excitation force and external sinusoïdal forced, we obtained  the parametric
 dissipative  modified Rayleigh-Duffing 
oscillator which is expressed by Eq. (\ref{eq.0}). We  will study the harmonic vibration,  the bifurcation and transition to chaos of the system.
The effects of the nonlinear damping, the parametric excitation amplitude and the external forcing amplitude will be seecked. Through this work, we will
find  our modified Rayleigh-Duffing oscillator regular and chaotic behaviors.e 

\section{ Harmonic oscillatory states}
Assuming that the fundamental component of the solution and the external 
excitation have the same period,  the amplitude of harmonic oscillations can be
tackled using the harmonic balance method \cite{2}. For this purpose,  we express
its solutions as
\begin{eqnarray}
 x &=& A \cos\left(\Omega-\psi\right)t  +\xi \label{eq1}
\end{eqnarray}
where $A$ represents the amplitude of the oscillations and $\xi$
a constant. 

Inserting this solution Eq.(\ref{eq1})in Eq.(\ref{eq.0})
 and equating the constants and the coefficients of
 $\sin{\Omega}t$ and $\cos{ \Omega t}$, we have

\begin{eqnarray}
 &&[-A\Omega^2-\frac{1}{2}\epsilon\beta\Omega^2 A^2+\alpha\xi+\gamma A+3\lambda\xi^2 A+\frac{3}{4}\lambda A^3+\frac{1}{4}\epsilon k_2 \Omega^2A^3]^2+\cr
[&&-\epsilon\mu\Omega A+\frac{3}{4}\mu\epsilon\Omega^3 A^3-\epsilon k_1\Omega\xi A]^2=F_0^2, \label{eq2}
\end{eqnarray}
\begin{equation}
 \frac{1}{2}\epsilon\beta\Omega^2A^2+\frac{1}{2}\epsilon k_2\xi\Omega^2A^2+\frac{1}{2}\alpha A+\gamma\xi+\lambda\xi^3+\frac{3}{2}\lambda\xi A^2=0. 
\label{eq3}
\end{equation}

 If it is assumed that $|{\xi}|\ll |A|$,  i.e that shift in $x = 0$ 
 is small compared to the amplitude \cite{8} ,  then ${\xi}^{2}$ 
and ${\xi}^{3}$  terms can be neglected,  Eq.(\ref{eq3}) become

\begin{equation}
 \frac{1}{2}\epsilon\beta\Omega^2A^2+\frac{1}{2}\epsilon k_2\xi\Omega^2A^2+\frac{1}{2}\alpha A+\gamma\xi+\frac{3}{2}\lambda\xi A^2=0. \label{eq4}
\end{equation}
We obtained
\begin{equation}
 \xi=\frac{\frac{1}{2}\epsilon\beta\Omega^2A^2+\frac{1}{2}\alpha A}{-\gamma-\frac{1}{2}(\epsilon k_2\Omega^2+3\lambda)A^2}. \label{eq5}
\end{equation}
Substituting Eq.(\ref{eq5})  into Eq. (\ref{eq2}) leads us to the following nonlinear algebraic equation
\begin{eqnarray}
 &&[-A\Omega^2-\frac{1}{2}\epsilon\beta\Omega^2 A^2+\gamma A+ A+\frac{1}{4}(\epsilon k_2 \Omega^2+3\lambda )A^3+\cr
&&\frac{\frac{1}{2}\epsilon\alpha\beta\Omega^2A^2+\frac{1}{2}\alpha^2 A}{-\gamma-\frac{1}{2}(\epsilon k_2\Omega^2+3\lambda)A^2}]^2+\cr
&&[-\epsilon\mu\Omega A+\frac{3}{4}\epsilon\mu\Omega^3 A^3-\epsilon k_1\Omega\frac{\frac{1}{2}\epsilon\beta\Omega^2A^2+\frac{1}{2}\alpha A}
{-\gamma-\frac{1}{2}(\epsilon k_2\Omega^2+3\lambda)A^2} A]^2=F_0^2, \label{eq6}
\end{eqnarray}
After some algebraics manipulations, Eq. (\ref{eq6}) can been rewritten as follow:

\begin{eqnarray}
&&(a^2+f^2)A^{10}+2abA^9+(b^2+2ac+2fg)A^8+(2bc+2ad+2fh)A^7+\cr
&&(c^2+2bd+2ae+2fi+g^2)A^6+(2ed+2be+2gh)A^5+\cr
&&(d^2+j+h^2+2ce+2gi)A^4+(2de+2hi)A^3+\cr
&&(e^2+i^2+k)A^2+l=0, \label{eq8}
\end{eqnarray}
with 
\begin{eqnarray}
 &&a=-\frac{1}{8}(\epsilon k_2\Omega^2+3\lambda)^2,\quad b=\frac{1}{4}\epsilon\beta\Omega^2(\epsilon k_2\Omega^2+3\lambda),\cr
&&c=-\frac{1}{4}(3\gamma-2\Omega^2)(\epsilon k_2\Omega^2+3\lambda),\quad d=\frac{1}{2}\epsilon\beta\Omega^2(\gamma+\alpha),\cr
&&e=\frac{1}{2}\alpha^2-\gamma(\gamma-\Omega^2), \quad f=-\frac{3}{8}\epsilon\mu\Omega^2(\epsilon k_2\Omega^2+3\lambda), \cr
&&g=\frac{1}{2}[\epsilon\mu\Omega(\epsilon k_2\Omega^2+3\lambda)-\frac{3}{2}\epsilon\mu\gamma\Omega^2-\epsilon^2\beta k_1\Omega^3],\cr
&&h=-\frac{1}{2}\epsilon^2\alpha k_1\Omega,\quad i=\epsilon\mu\gamma\Omega, \quad j=-\frac{1}{4}F_0^2(\epsilon k_2\Omega^2+3\lambda)^2,\cr
&&k=\gamma F_0^2(\epsilon k_2\Omega^2+3\lambda), \quad l=-F_0^2\gamma^2.\label{eq9}
\end{eqnarray}

\begin{figure}[htbp]
\begin{center}
 \includegraphics[width=8cm,  height=8cm]{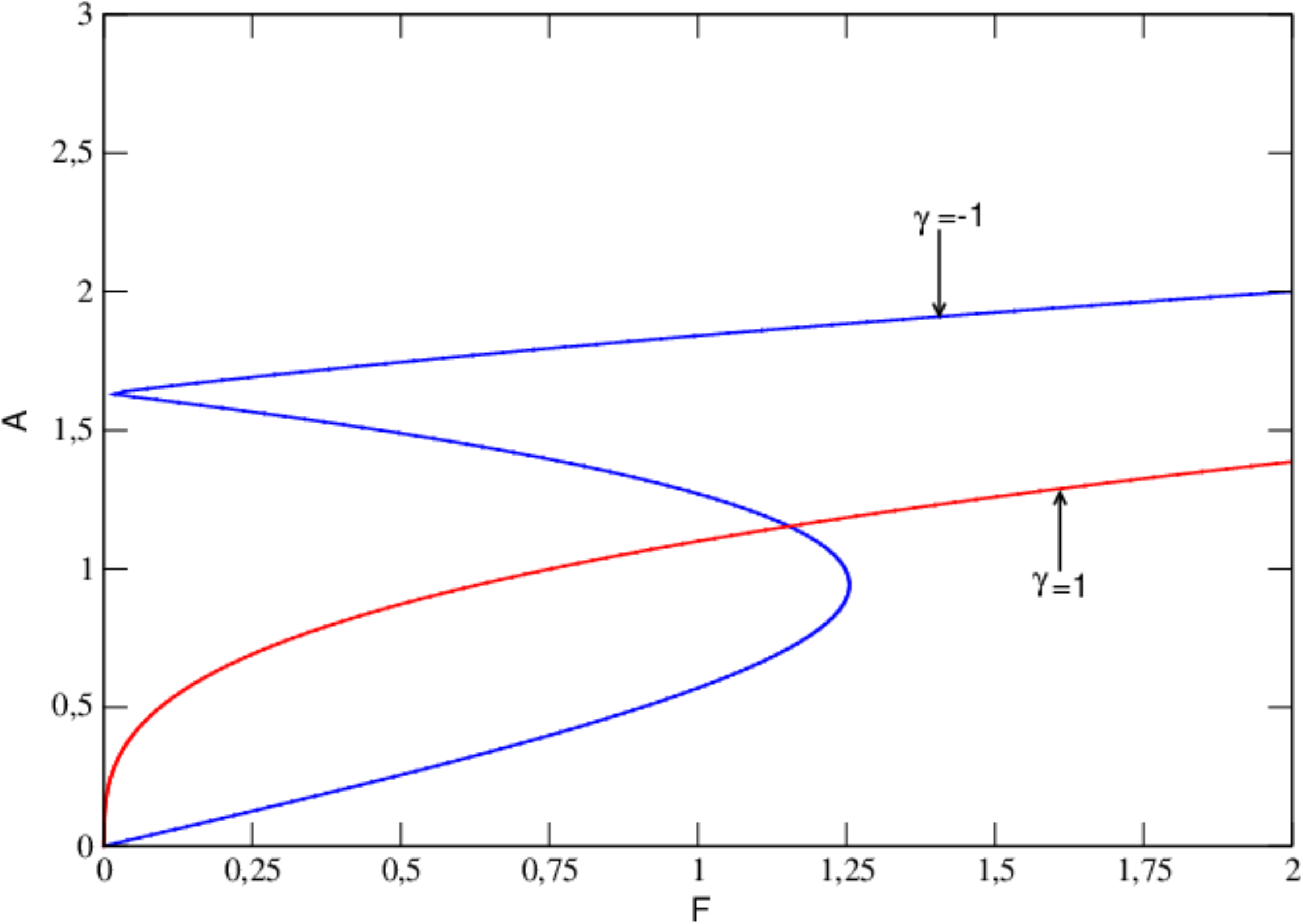}
\end{center}
\caption{Effects of $\gamma $ on the amplitude-response curves with  the parameters $\alpha=0, \beta=0.5,k_1=0.5, k_2=0.5,
 \mu=0.5,  \lambda=1$ and $ \Omega=1$.}
\label{fig:1}
\end{figure}

\begin{figure}[htbp]
\begin{center}
 \includegraphics[width=12cm,  height=6cm]{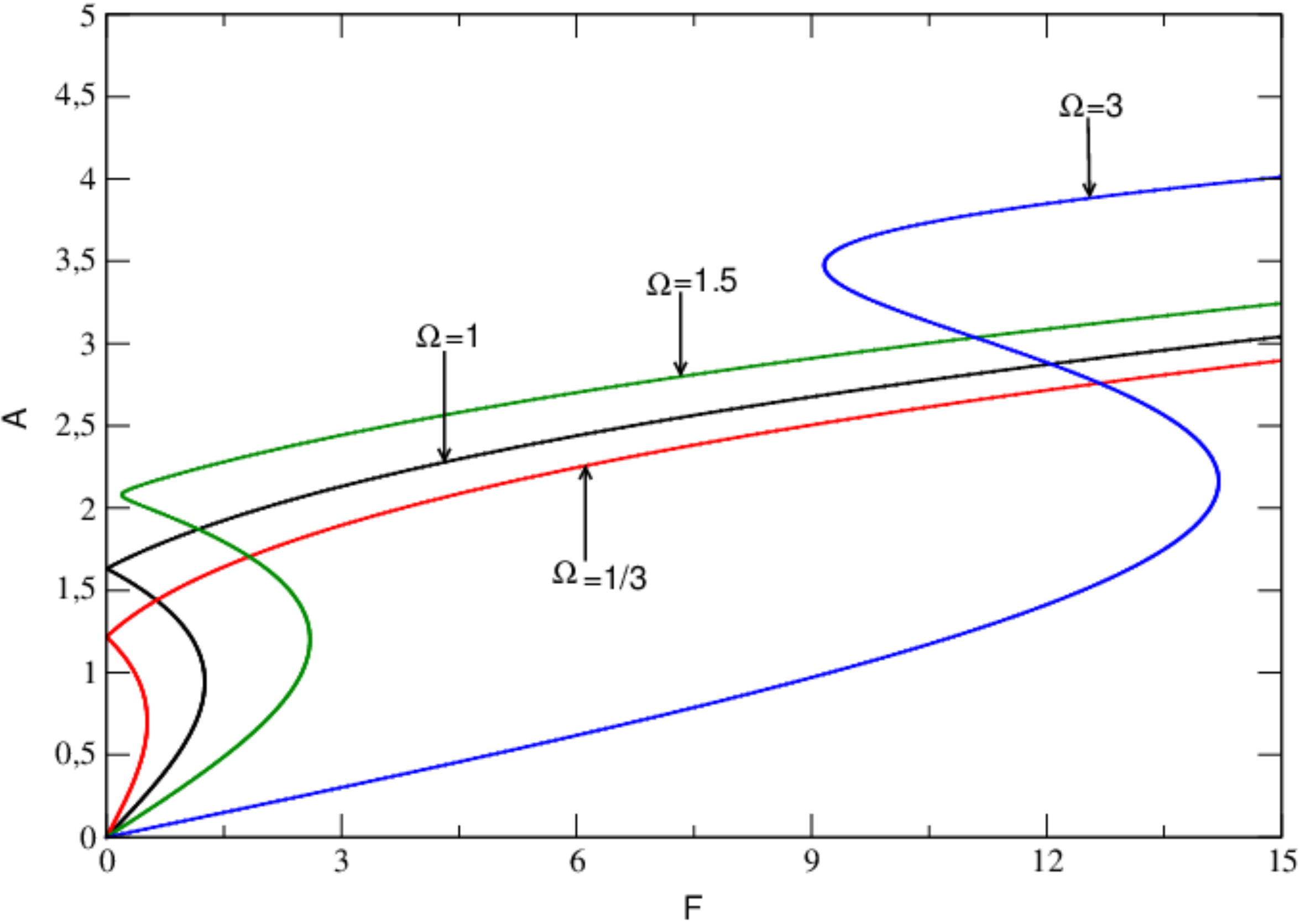}
\end{center}
\caption{Effects of $\Omega$ on the amplitude-response curves with $ \gamma=-1$, and the parameters of Fig.\ref{fig:1}.}
\label{fig:2}
\end{figure}

\begin{figure}[htbp]
\begin{center}
 \includegraphics[width=12cm,  height=6cm]{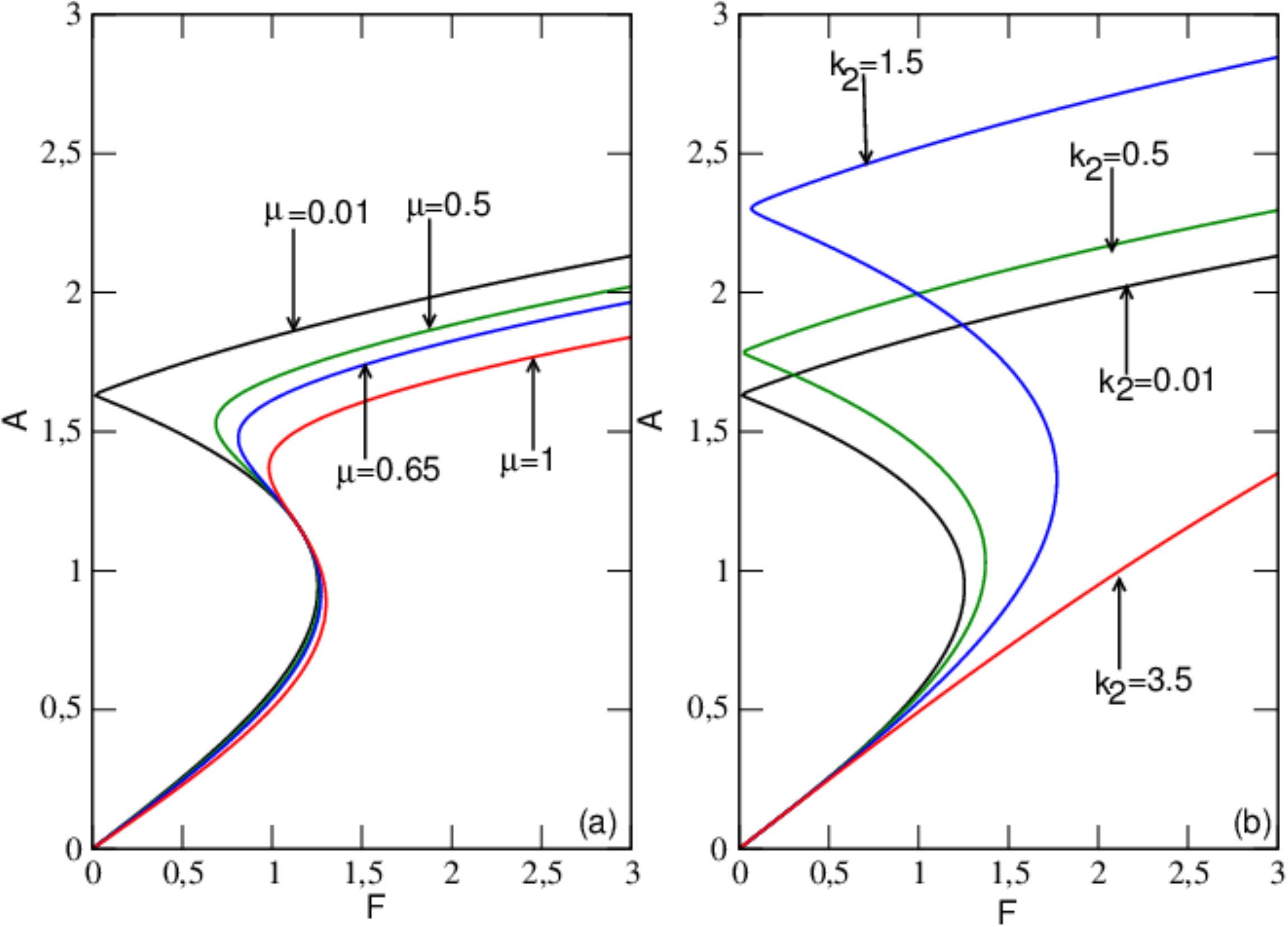}
\includegraphics[width=12cm,  height=6cm]{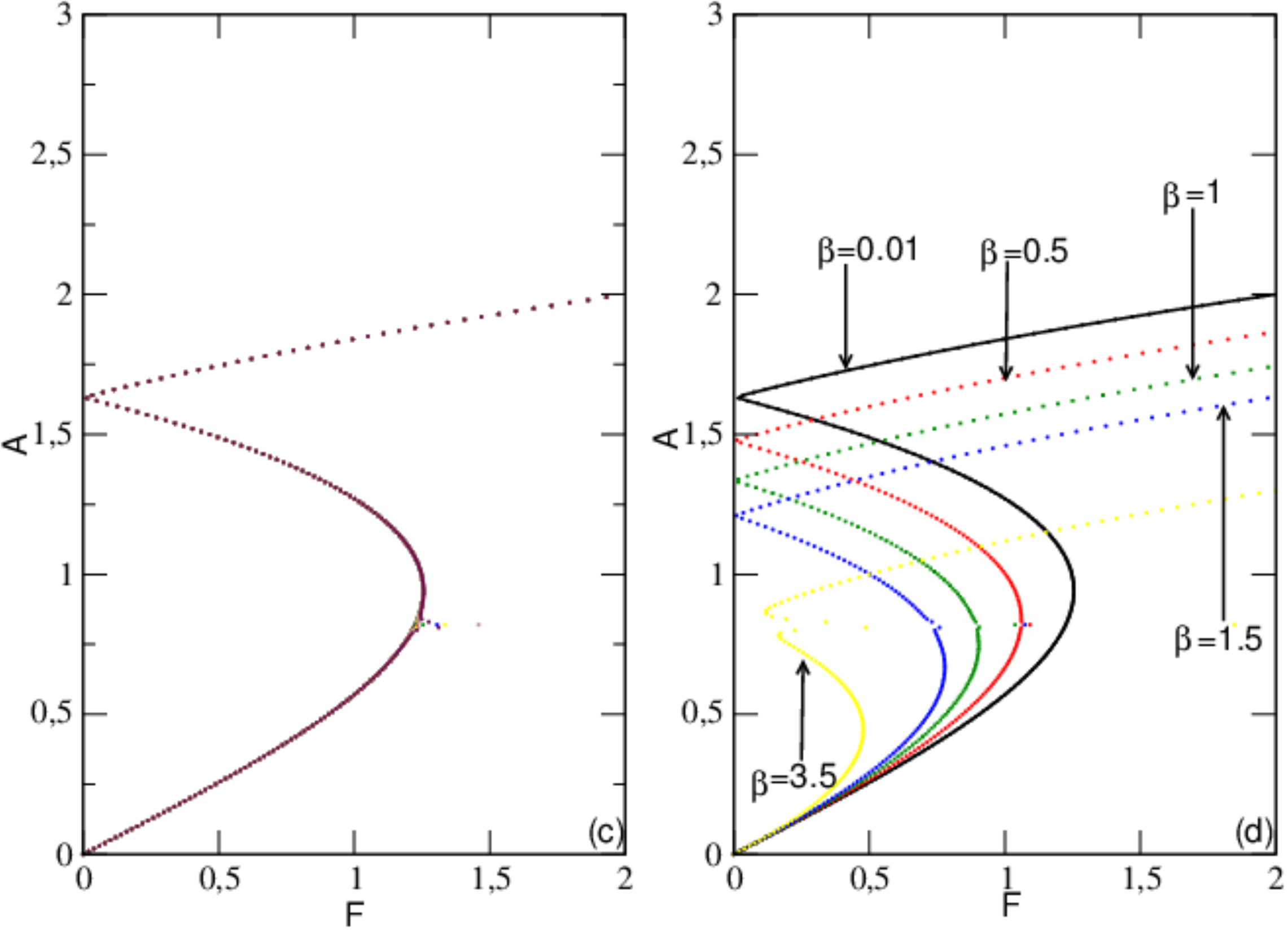}
\includegraphics[width=12cm,  height=6cm]{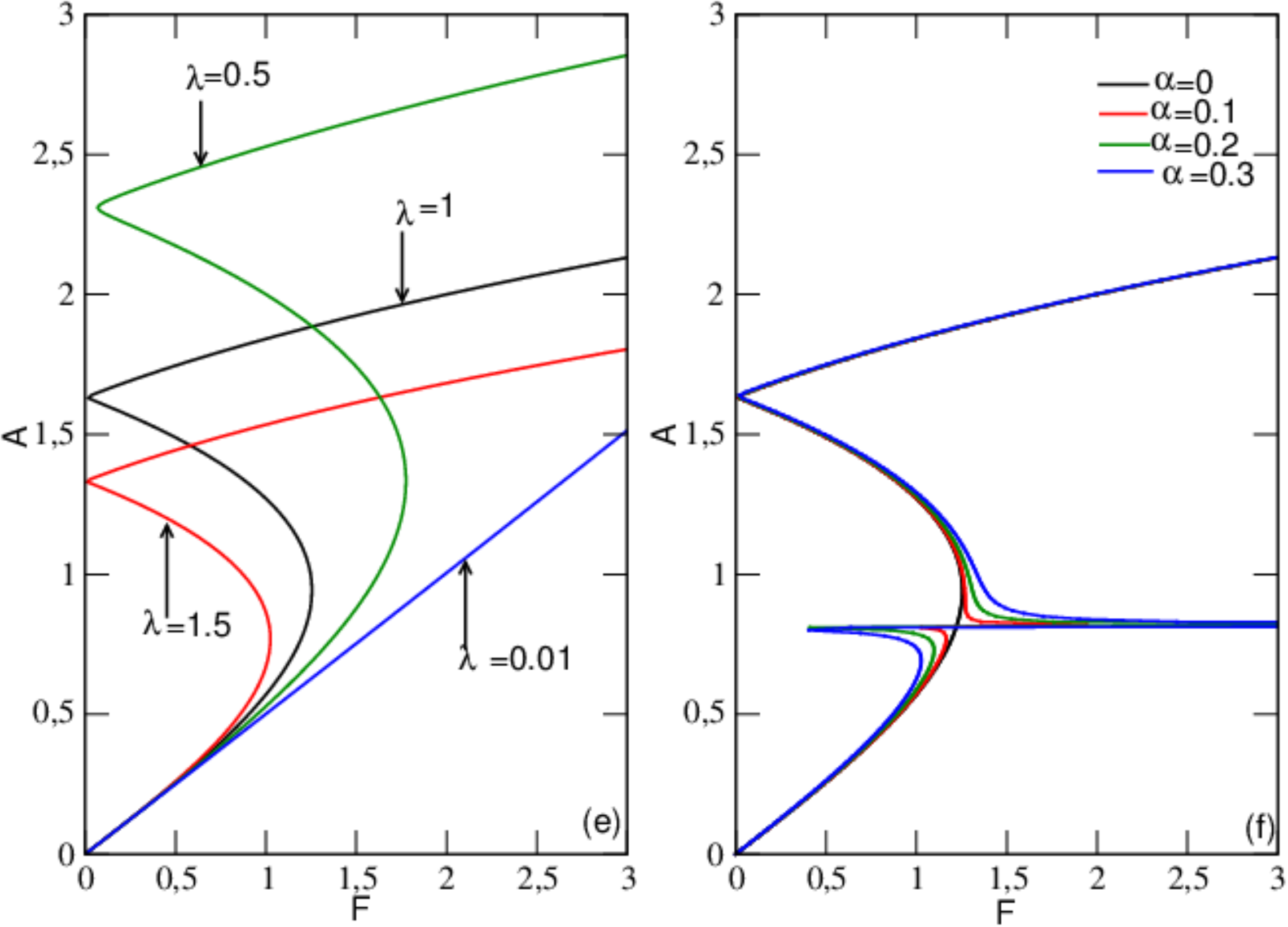}
\end{center}
\caption{$(a):$ Effect of $\mu$, $(b):$ effect of $k_2$, $(c):$ effect of $k_1$, $(d):$ effect of $\beta$, $(d):$ effect of $\lambda$ and
 $(e):$ effect of $\alpha$ on the amplitude-response curves with $ \Omega=1$, and the parameters of Fig. \ref{fig:1}.}
\label{fig:3}
\end{figure}

\begin{figure}[htbp]
\begin{center}
 \includegraphics[width=8cm,  height=8cm]{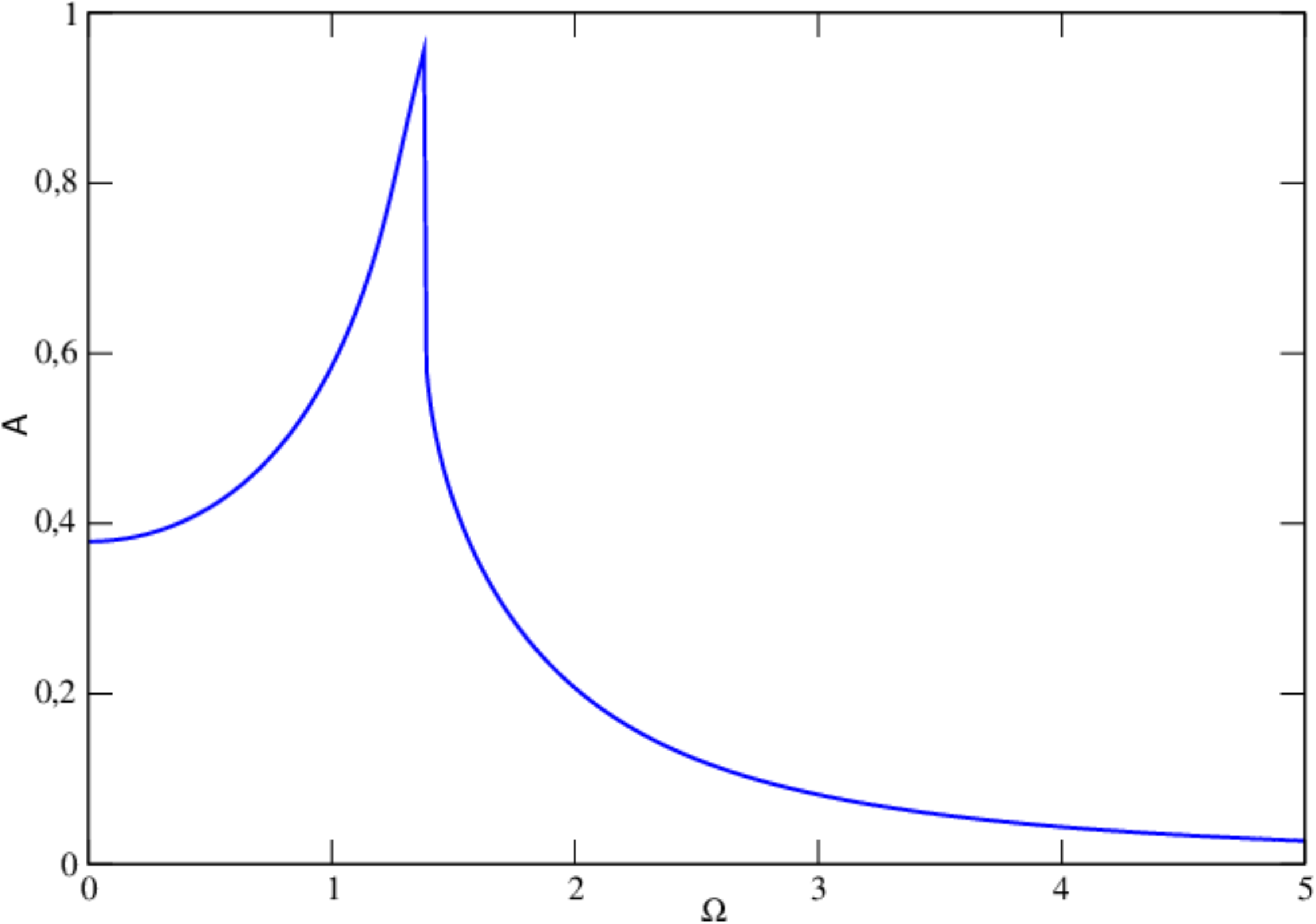}
\end{center}
\caption{ Frequency-response curves with the parameters $\alpha=0, \beta=0.5, k_1=0.5, k_2=0.5,
 \mu=0.25, \lambda=1, \gamma=1$ and $F=0.65$.}
\label{fig:4}
\end{figure}

\begin{figure}[htbp]
\begin{center}
 \includegraphics[width=12cm,  height=10cm]{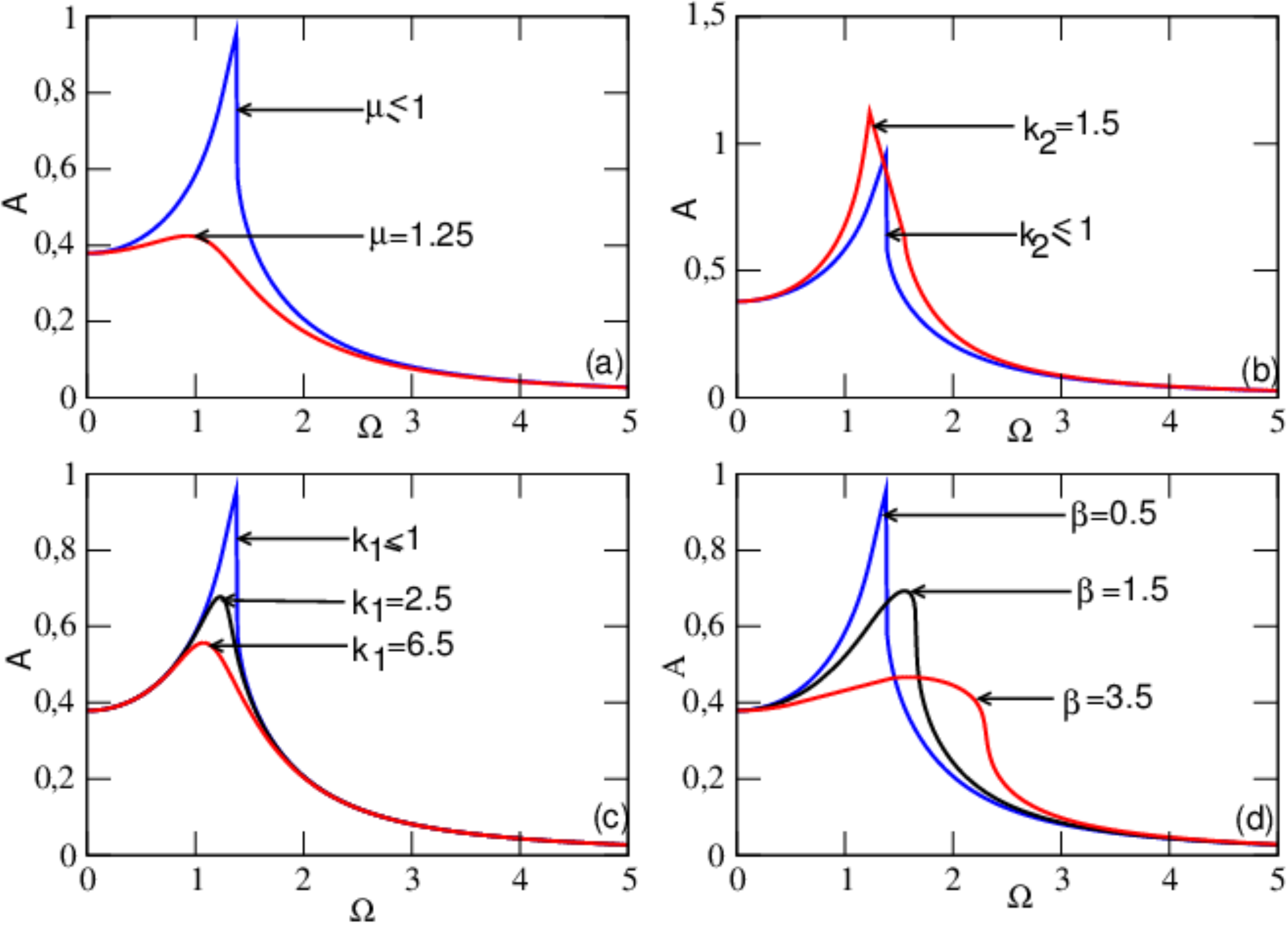}
\end{center}
\caption{ Effects of damping on frequency-response curves with the parameters of Fig.\ref{fig:6} $(a)$ effect of $\mu$, $(b)$ effect of $k_2$, 
$(c)$: effect of $k_1$ and $(d)$ effect of $\beta$.}
\label{fig:5}
\end{figure}

\begin{figure}[htbp]
\begin{center}
 \includegraphics[width=12cm,  height=8cm]{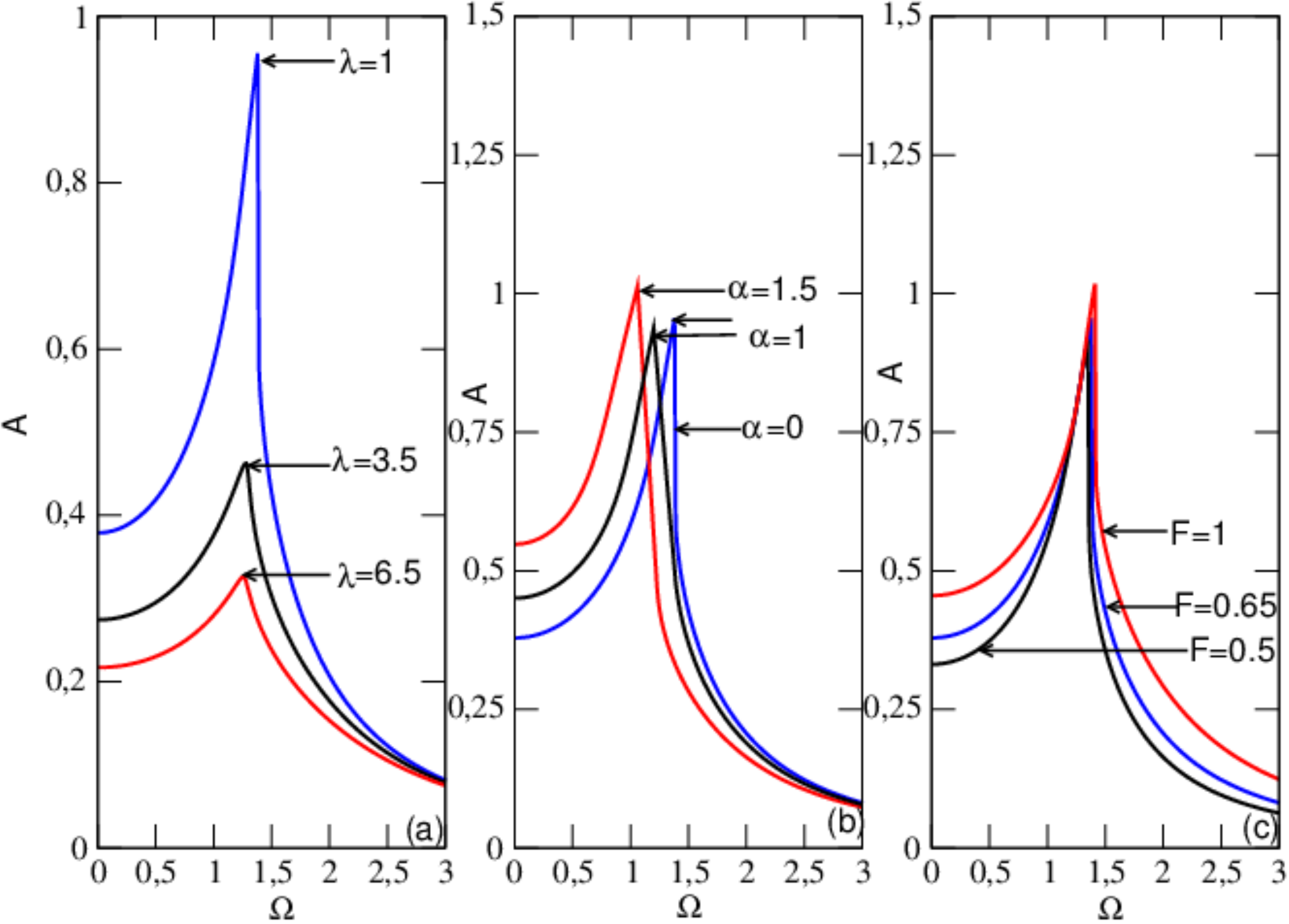}
\end{center}
\caption{ Effects of other  on parameters frequency-response curves with the parameters of Fig.\ref{fig:6} $(a)$ effect of $\lambda$, $(b)$ effect of $\alpha$, 
$(c)$: effect of amplitude of external forced $F$.}
\label{fig:6}
\end{figure}

We investigate the effects of system parameters on the amplitude of oscillations $A$ by solving Eq. (\ref{eq8}) using the Newton-Raphson algorithm.
The five first figures are  the amplitude-response curves and its showing the hysteresis and jump phenomena. Fig.\ref{fig:1} show the effect of $\gamma$
on amplitude-response curve where hysteresis and jump phenomena  appeared when $\gamma=-1$ and desappeared when $\gamma=1$. Fig.\ref{fig:2} illustrate
the effect of external frequency on amplitude-response curve. Through this figure we notice that the hysteresis phenomenon desappeared when the external 
frequency increasing.  Figs.\ref{fig:3} $(a), (b), (c)$ and $(d)$ illustrate the effects of damping cofficients $\mu, k_2, k_1$ and $\beta$ respectively
 while the effects of the cubic nonlinear Duffing coefficient and amplitude of parametric excitation  are shown in Figs.\ref{fig:3} $(e)$ and $(f)$ 
respectively. Through these figures we noticed that when the cubic nonlinear damping cofficients $\mu$ and $k_2$ increasing, the hysteresis and jump 
phenomena disappear (see Figs. \ref{fig:3} $(a), (b)$) while these two phenomena became more several but amplitude of harmonic oscillations descrases
 when the pure quadratic nonlinear damping coefficient is increased (see. We notice that \ref{fig:3} $(c)$). The effect of unpure nonlinear quadratic 
damping parameter $k_1$ is not significated (see \ref{fig:3} $(d)$). One can notice that as $\lambda$ discreases, the 
jump and hysteresis phenomenon disappear (see \ref{fig:3} $(f)$). The topology of jump and hysteresis phenomenon curves
 is well influenced by the parametric excitation amplitude as shown in Fig. \ref{fig:3} $(a)$. In that figure, the amplitude-response curves are plotted
for different value of $\alpha$. One notices that as increases,  the topology of these two phenomenon are very modified. 
The behavior of the amplitude of this system oscillations is investigated when the external frequency $\Omega$
varies and the results are plotted in Fig. \ref{fig:4}  where  analytical resonance curves $A(\Omega)$ of the model is shown.
 The resonance obtained from Fig.\ref{fig:4}  is also affected by the nonlinear damping parameters, parametric excitation amplitude and external forced
amplitude (see Figs. \ref{fig:5}, \ref{fig:6}). Thereby, the observed resonant state obtained for a set of parameters can be destroyed according to the
 value taken by these parameters. For instance, when the damping cofficients $\mu, k_1, \beta $ increasing the frequency-response curve is destroyed, 
the peak value of resonance amplitude is descrased and we noticed that the resonance desappear (see Figs \ref{fig:5} $(a), (c),(d)$). Fig.\ref{fig:4} $(b)$
show that when $k_2$ is increased  the peak value of resonance amplitude incrased while the resonance frequency descreased. Through 
Figs. \ref{fig:6} $(a),(b),(c)$ we noticed also that the parameter $\lambda$ have the same effect that $\mu$ and $k_1$ on the frequency-response curve 
(see Fig.\ref{fig:6} $(a)$) while when $\alpha$ increasing the resonance frequency is descreased (see \ref{fig:6} $(b)$) and the frequency-response curve
become largest when the external forced amplitude is increased but the resonance frequency is not affected (see \ref{fig:6} $(c)$). 

\newpage
\section{Bifurcation and transition to chaos}

\begin{figure}[htbp]
\begin{center}
 \includegraphics[width=12cm,  height=4cm]{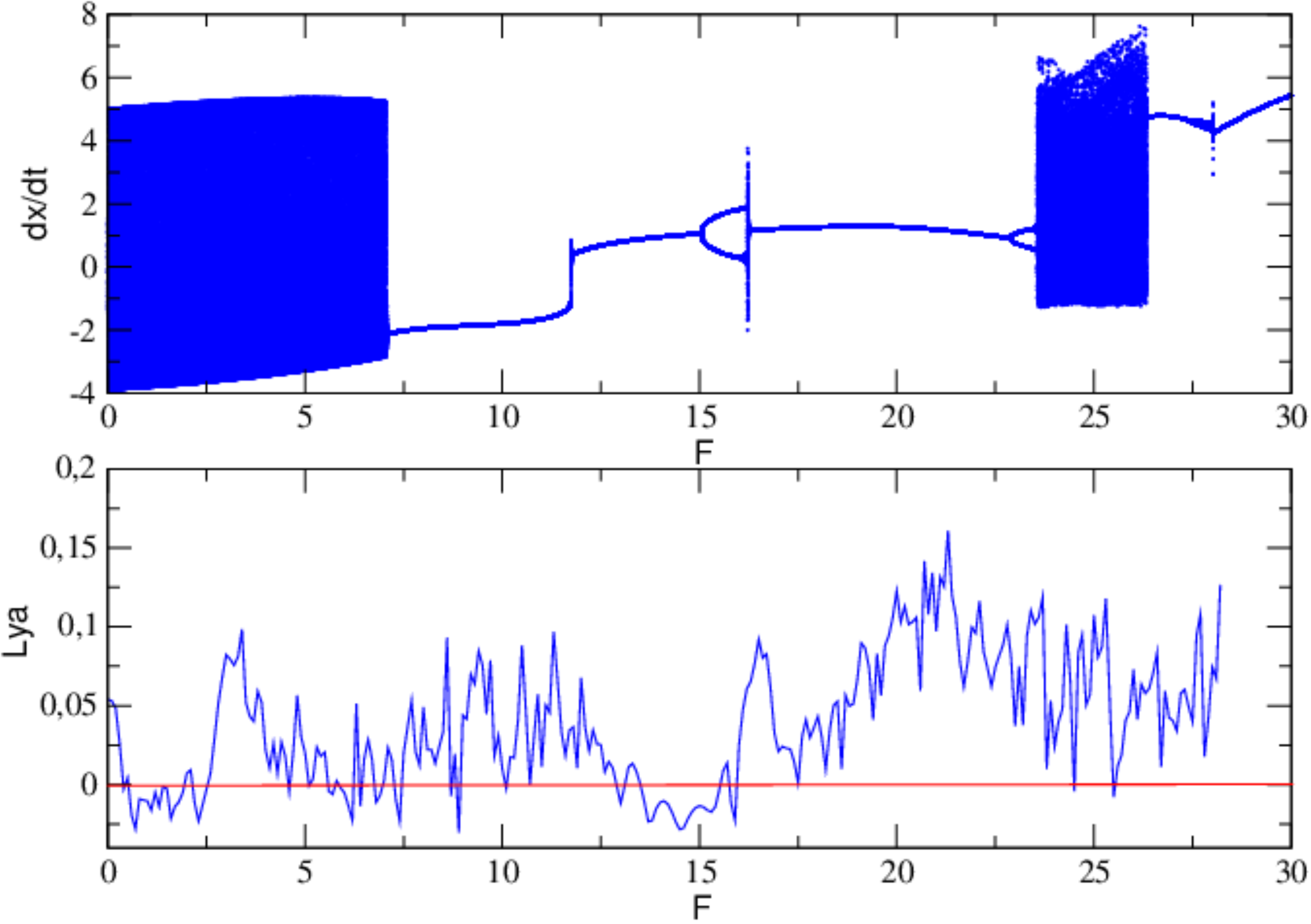}
\end{center}
\caption{Bifurcation diagram (upper frame) and Lyapunov exponent (lower frame) versus the amplitude F with parameters
 for $\alpha=0.3, \beta=0.05, k_1=0.05, k_2=0.05,
 \mu=0.0001, \lambda=1, \gamma=1$ and $\Omega=1$}
\label{fig:7}
\end{figure}

\begin{figure}[htbp]
\begin{center}
 \includegraphics[width=12cm,  height=4cm]{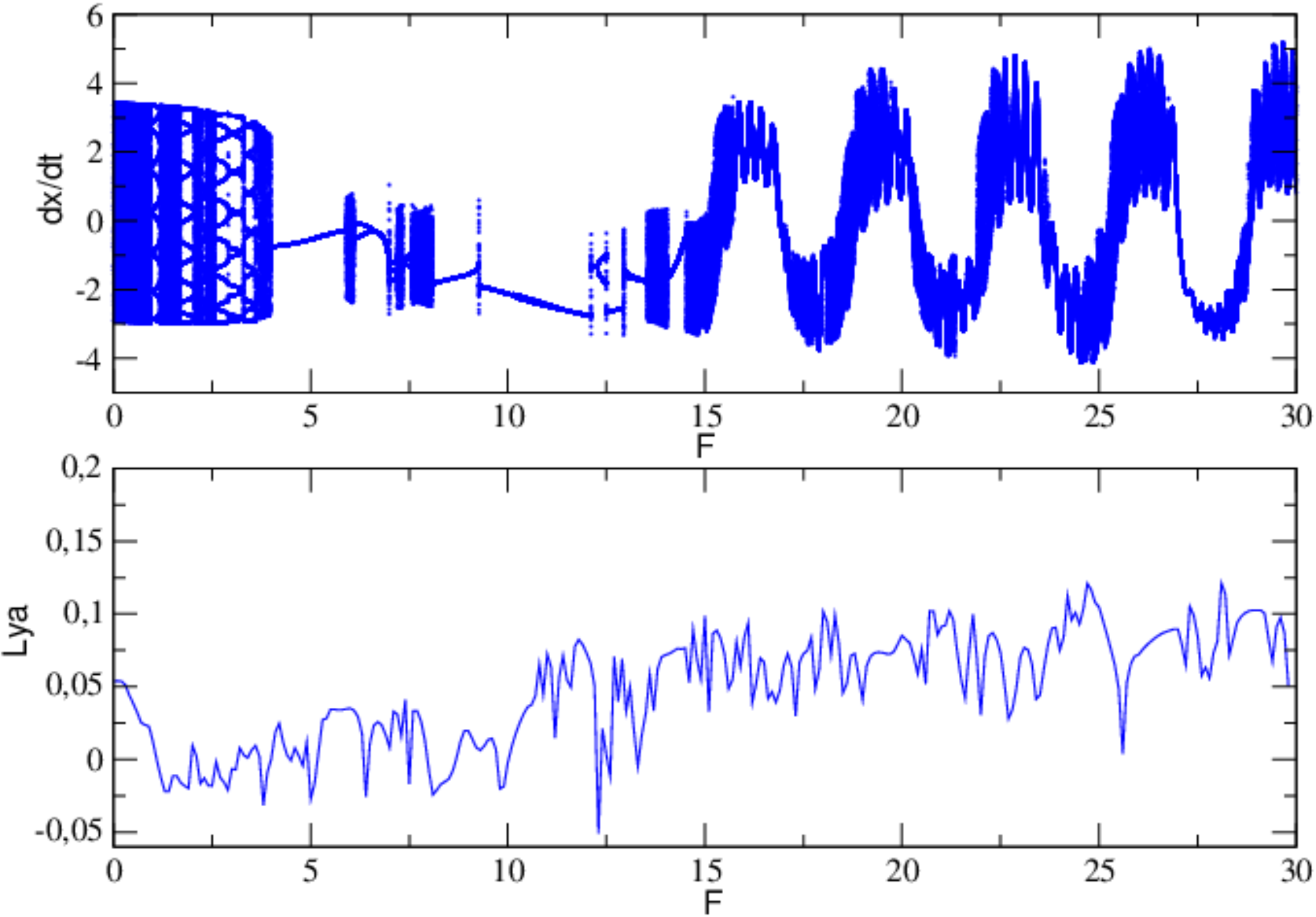}
\end{center}
\caption{Bifurcation diagram (upper frame) and Lyapunov exponent (lower frame) versus the amplitude F with parameters $\alpha=0.3, \beta=0.05, k_1=0.05, k_2=0.05,
 \mu=0.0001, \lambda=1, \gamma=1$ and $\Omega=1/3$}
\label{fig:8}
\end{figure}

\begin{figure}[htbp]
\begin{center}
 \includegraphics[width=12cm,  height=4cm]{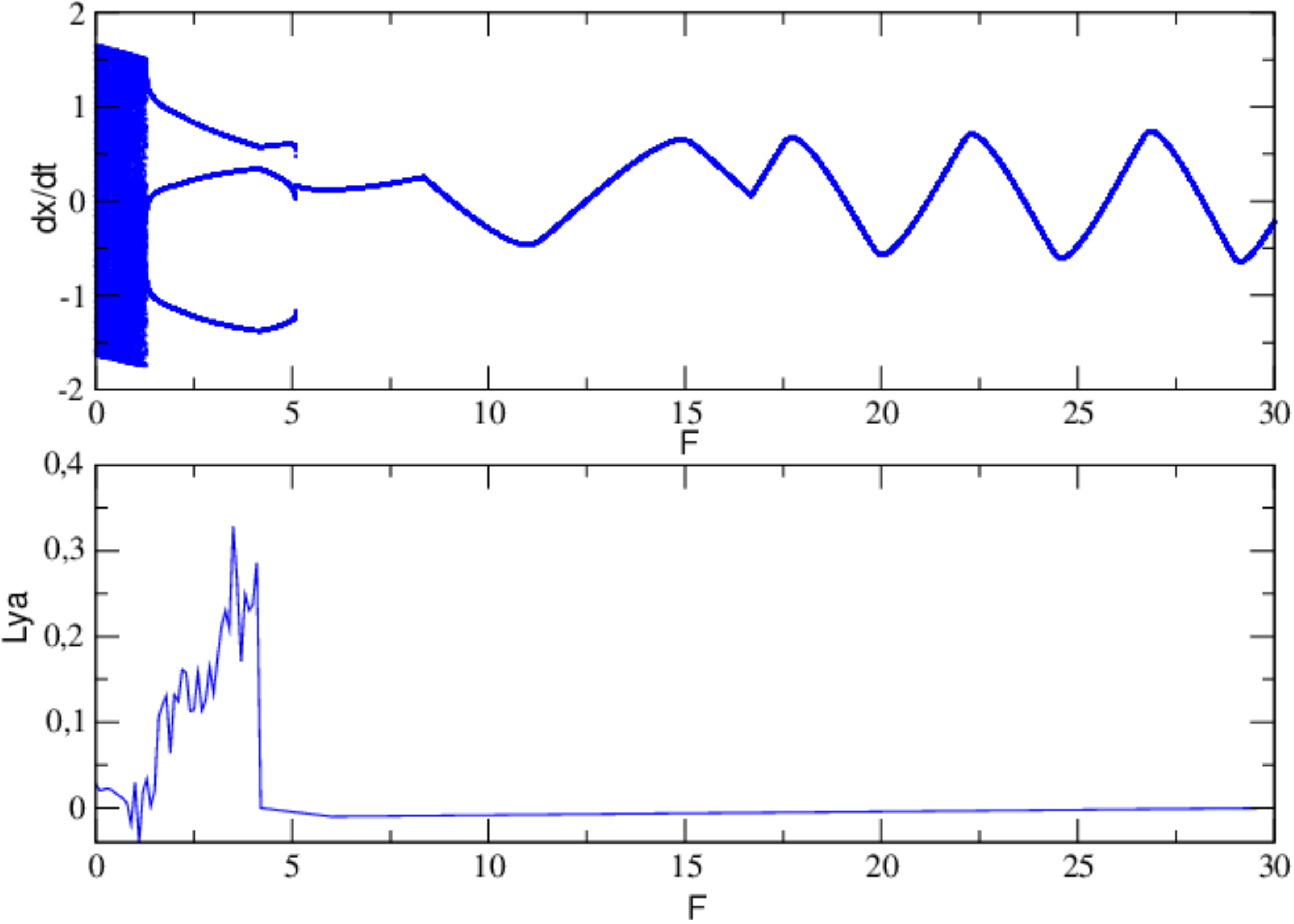}
\end{center}
\caption{Bifurcation diagram (upper frame) and Lyapunov exponent (lower frame) versus the amplitude F with parameters
 $\alpha=0.3, \beta=0.5, k_1=0.5, k_2=0.5,
 \mu=0.0001, \lambda=1, \gamma=1$ and $\Omega=3$}
\label{fig:9}
\end{figure}

\begin{figure}[htbp]
\begin{center}
 \includegraphics[width=12cm,  height=4cm]{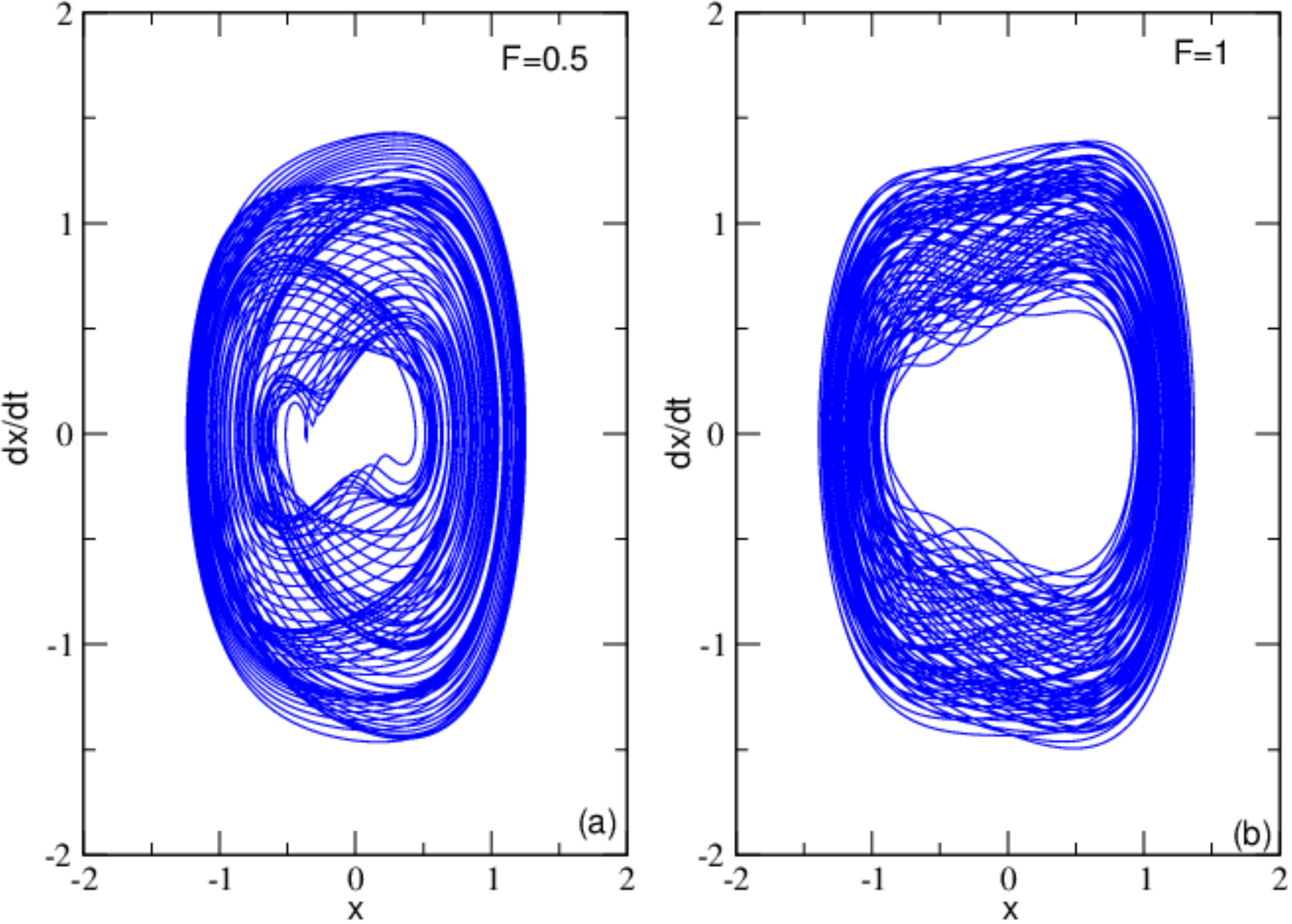}
\includegraphics[width=12cm,  height=4cm]{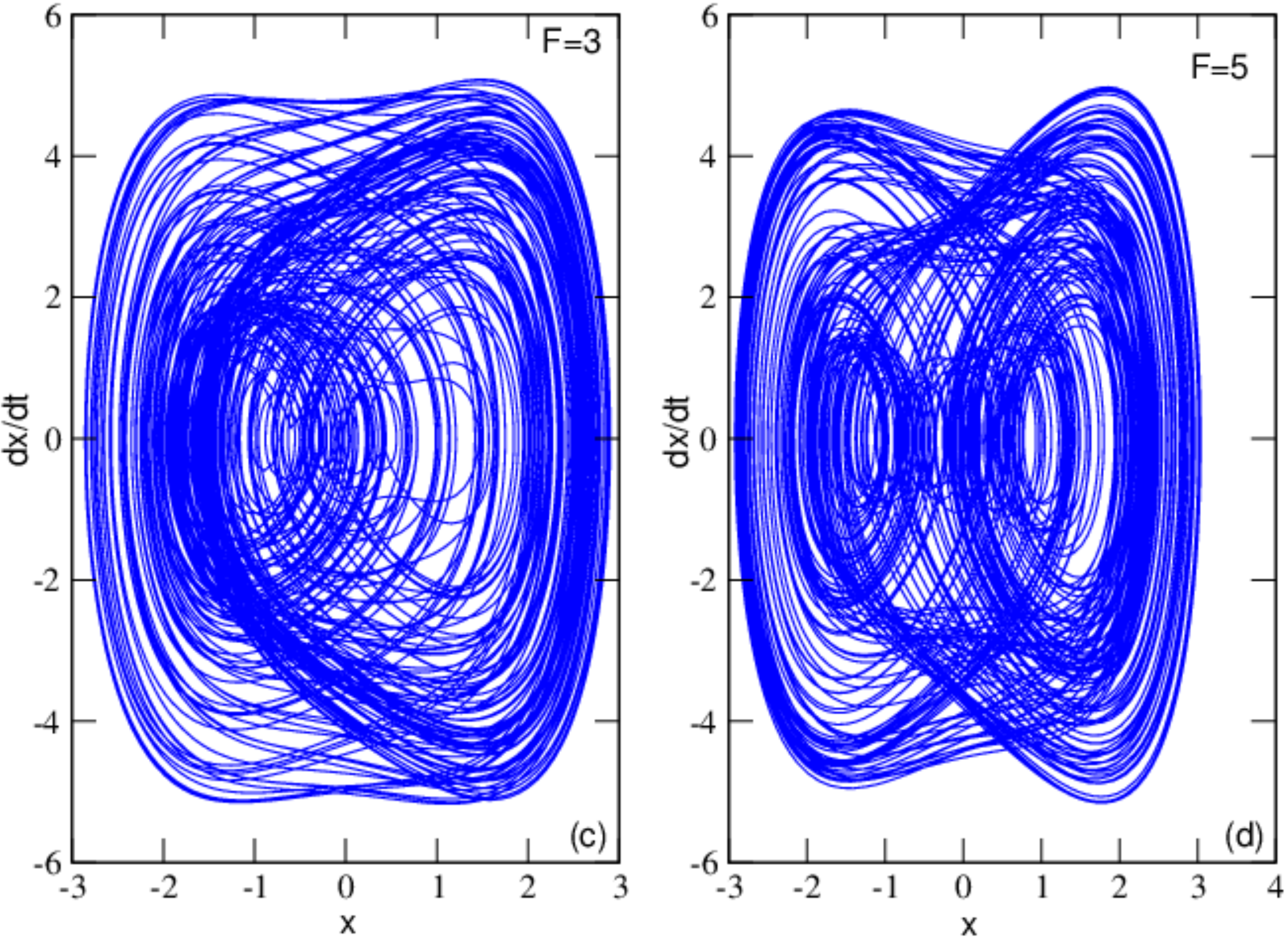}
\includegraphics[width=12cm,  height=4cm]{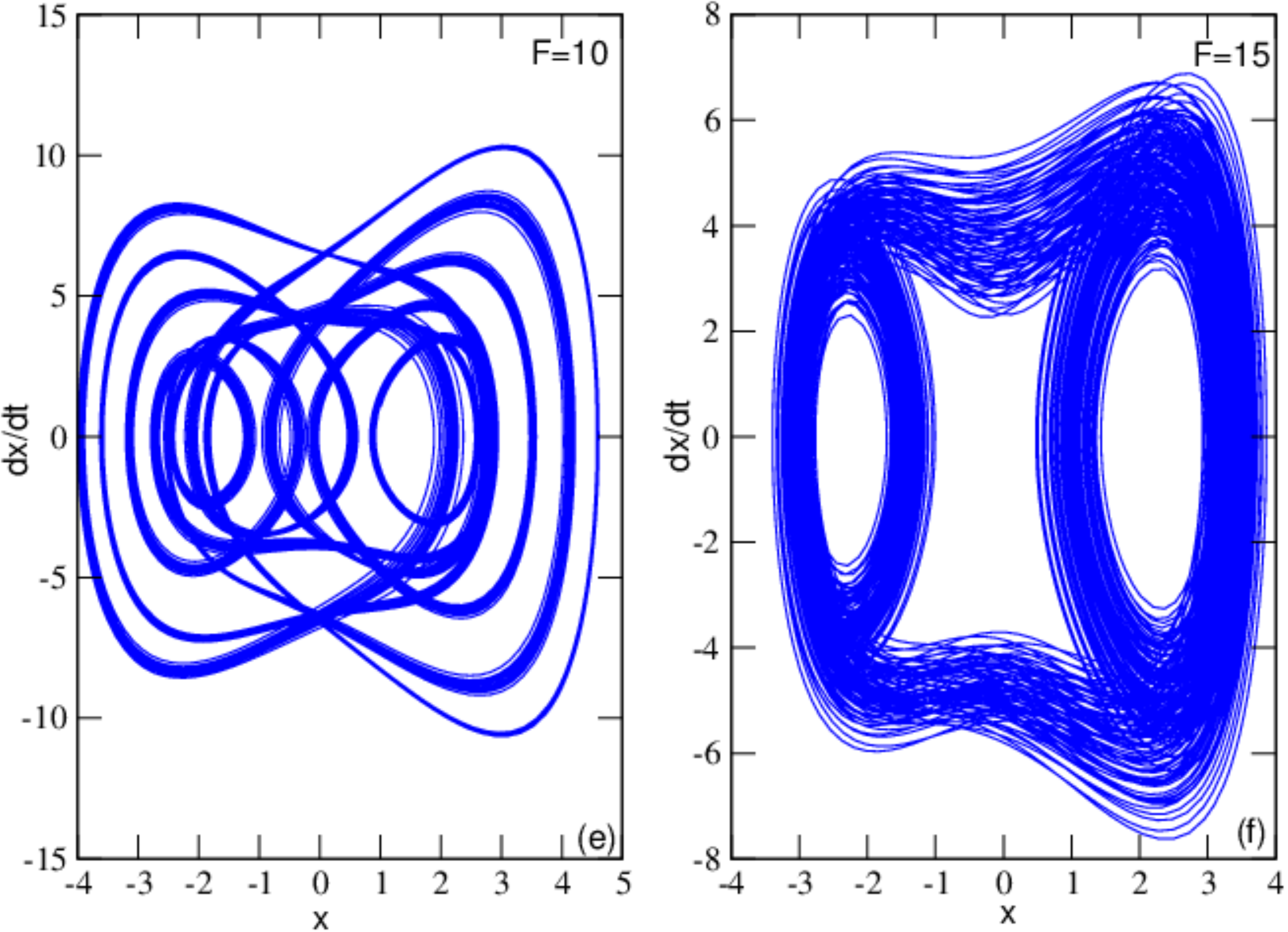}
\includegraphics[width=12cm,  height=4cm]{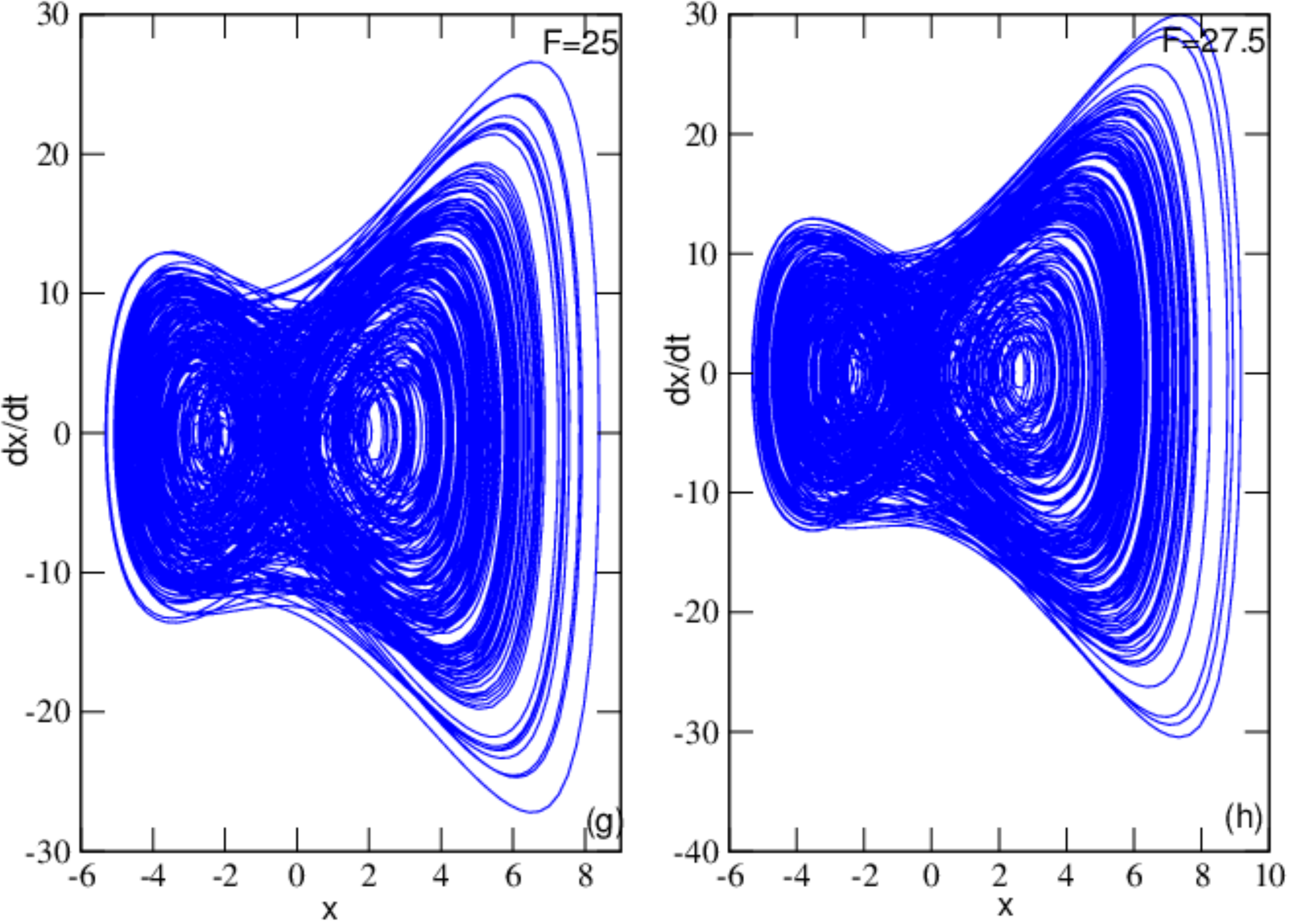}
\end{center}
\caption{Various phase portraits for several different values of F with the
parameters of Fig. \ref{fig:7}.}
\label{fig:10}
\end{figure}

\begin{figure}[htbp]
\begin{center}
 \includegraphics[width=12cm,  height=4cm]{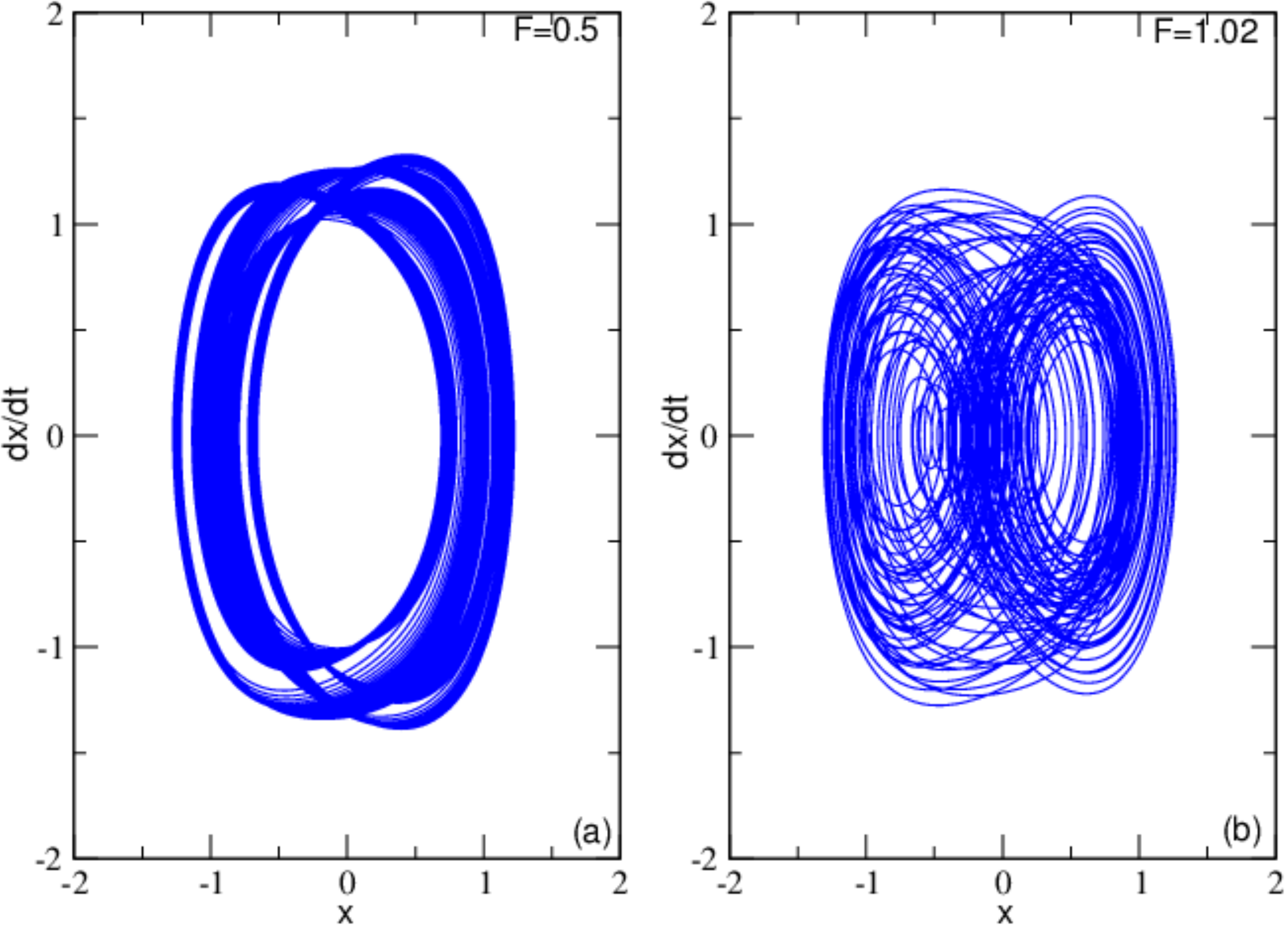}
\includegraphics[width=12cm,  height=4cm]{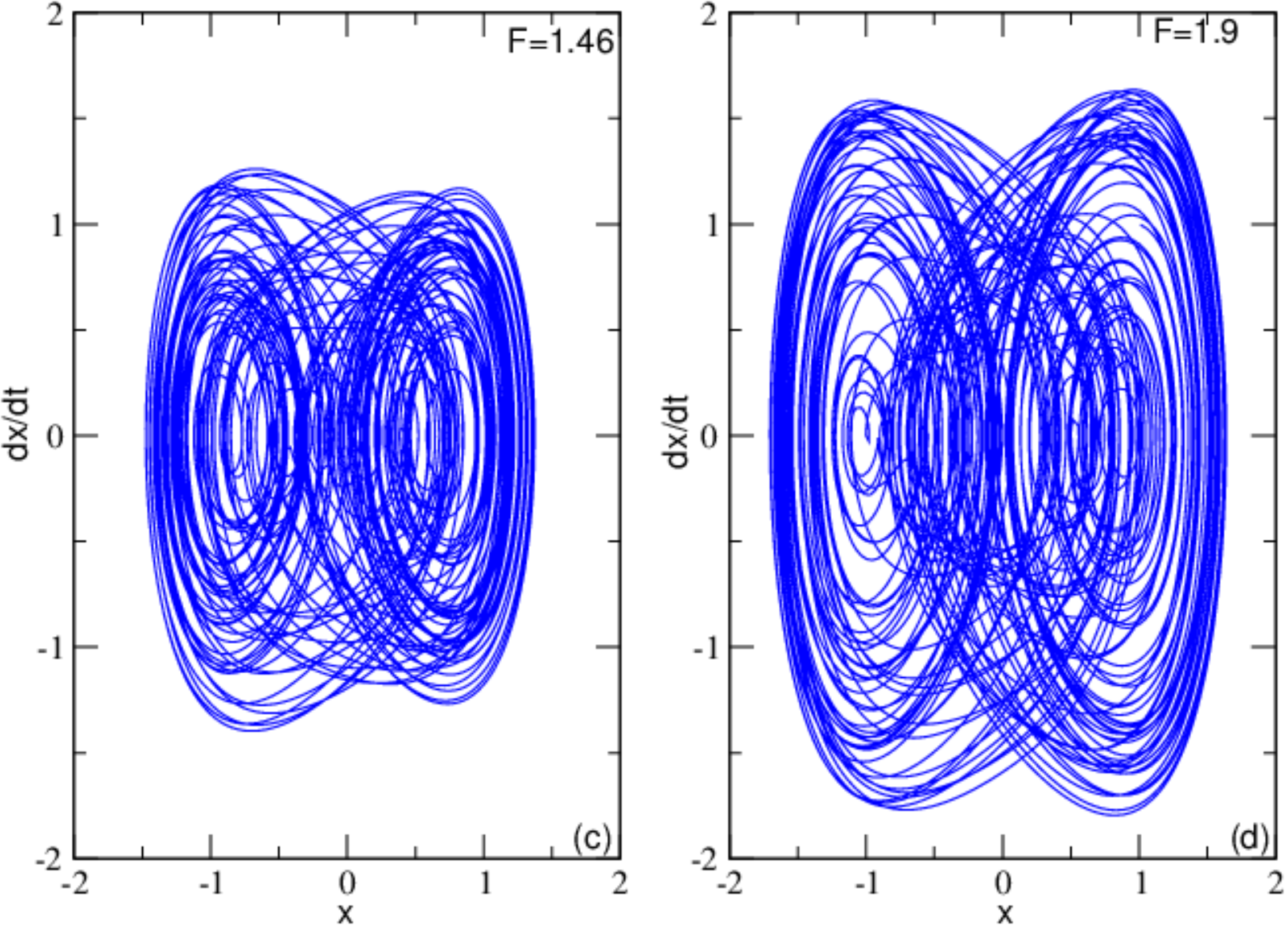}
\includegraphics[width=12cm,  height=4cm]{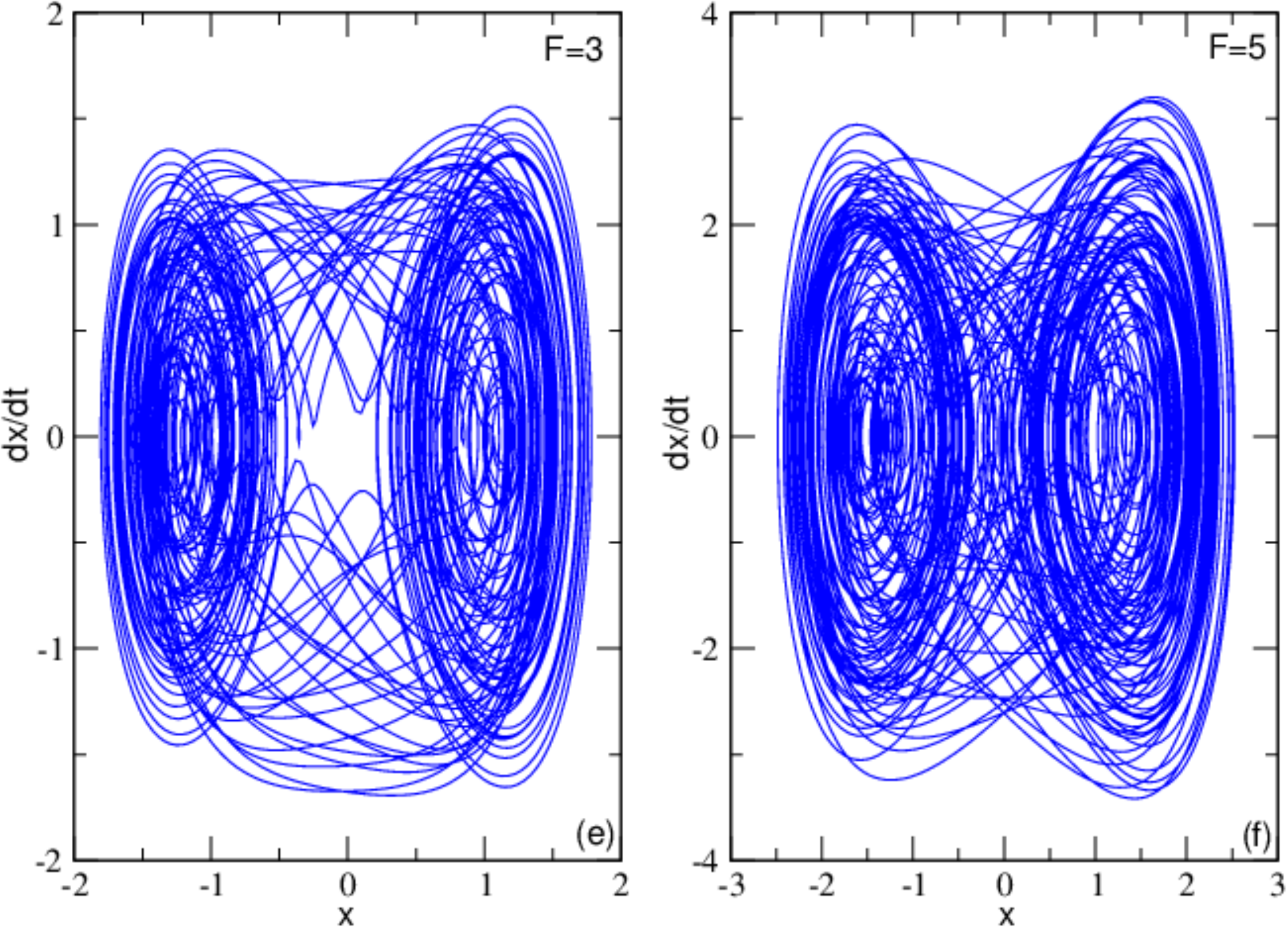}
\includegraphics[width=12cm,  height=4cm]{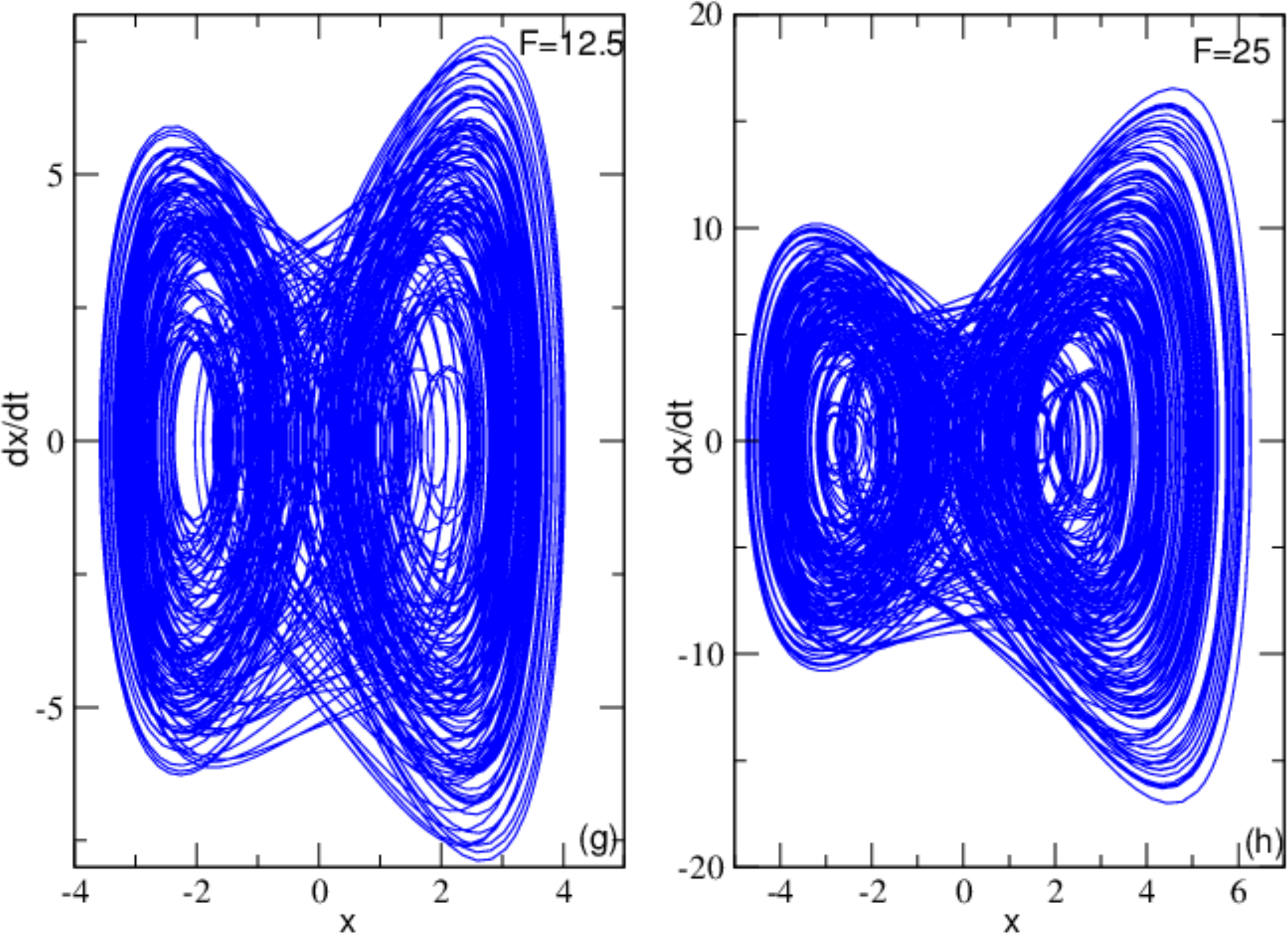}
\end{center}
\caption{Various phase portraits for several different values of F with the
parameters of Fig.\ref{fig:8}.}
\label{fig:11}
\end{figure}

\begin{figure}[htbp]
\begin{center}
 \includegraphics[width=12cm,  height=5cm]{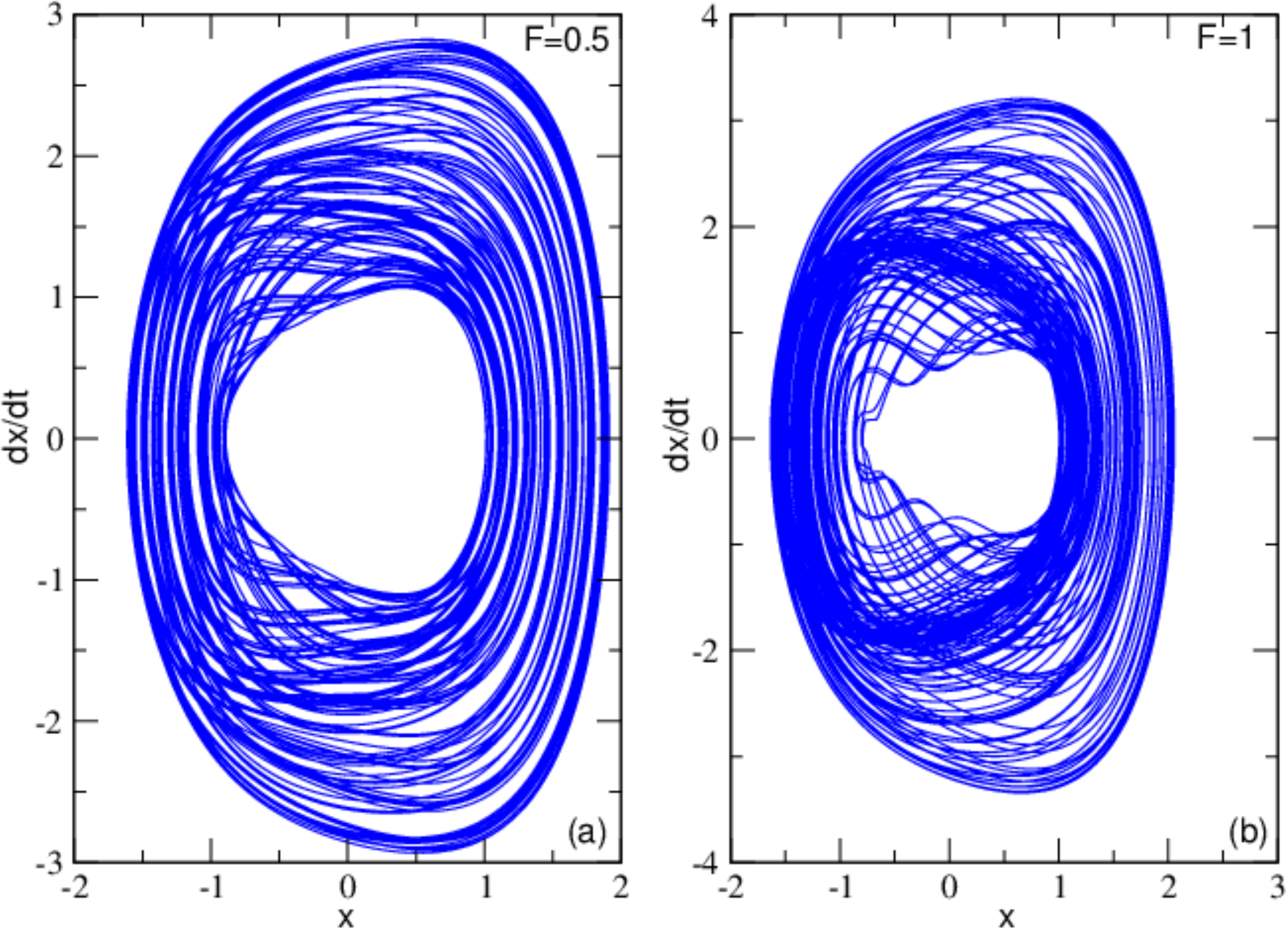}
\includegraphics[width=12cm,  height=5cm]{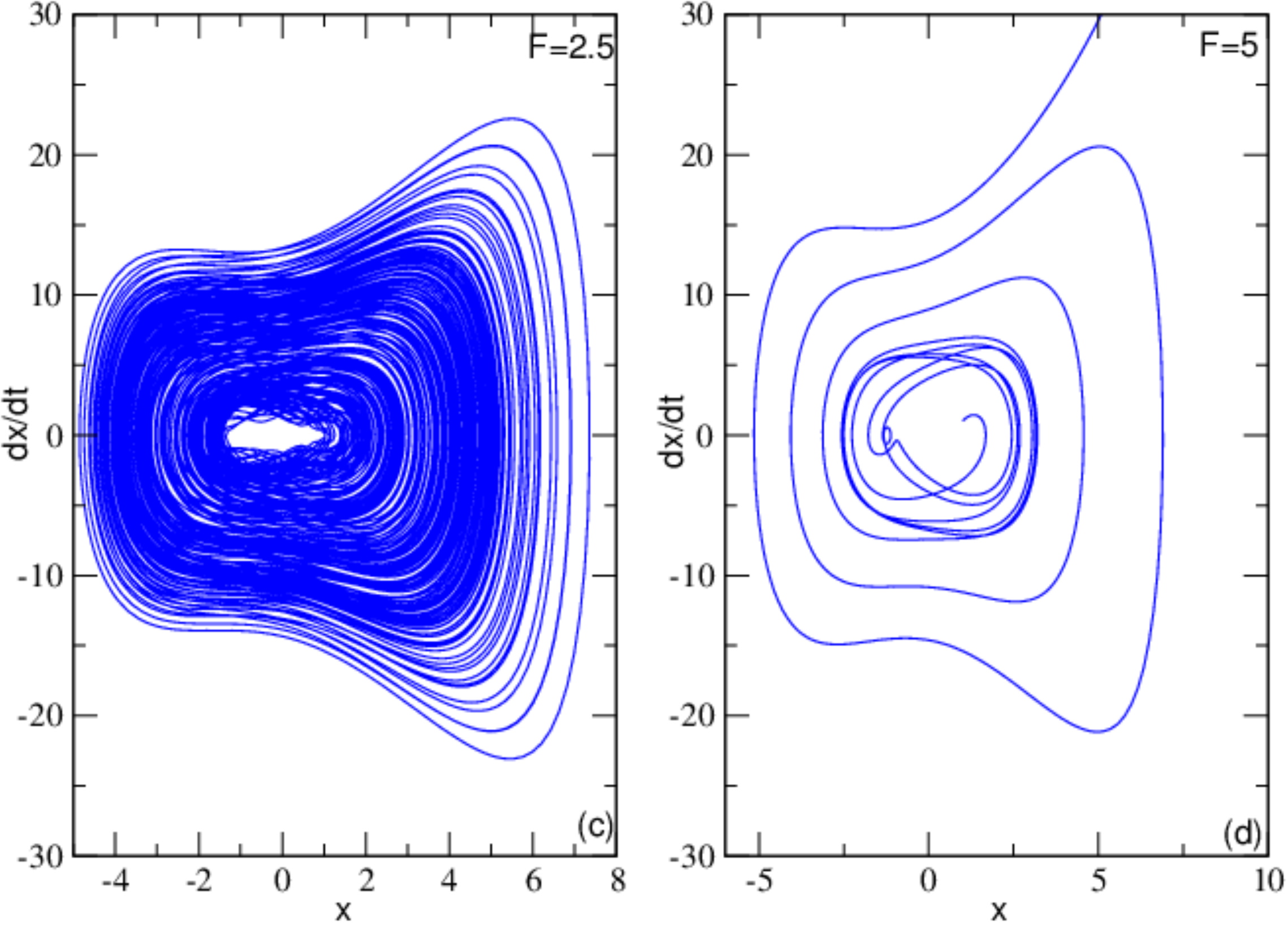}
\includegraphics[width=6cm,  height=6cm]{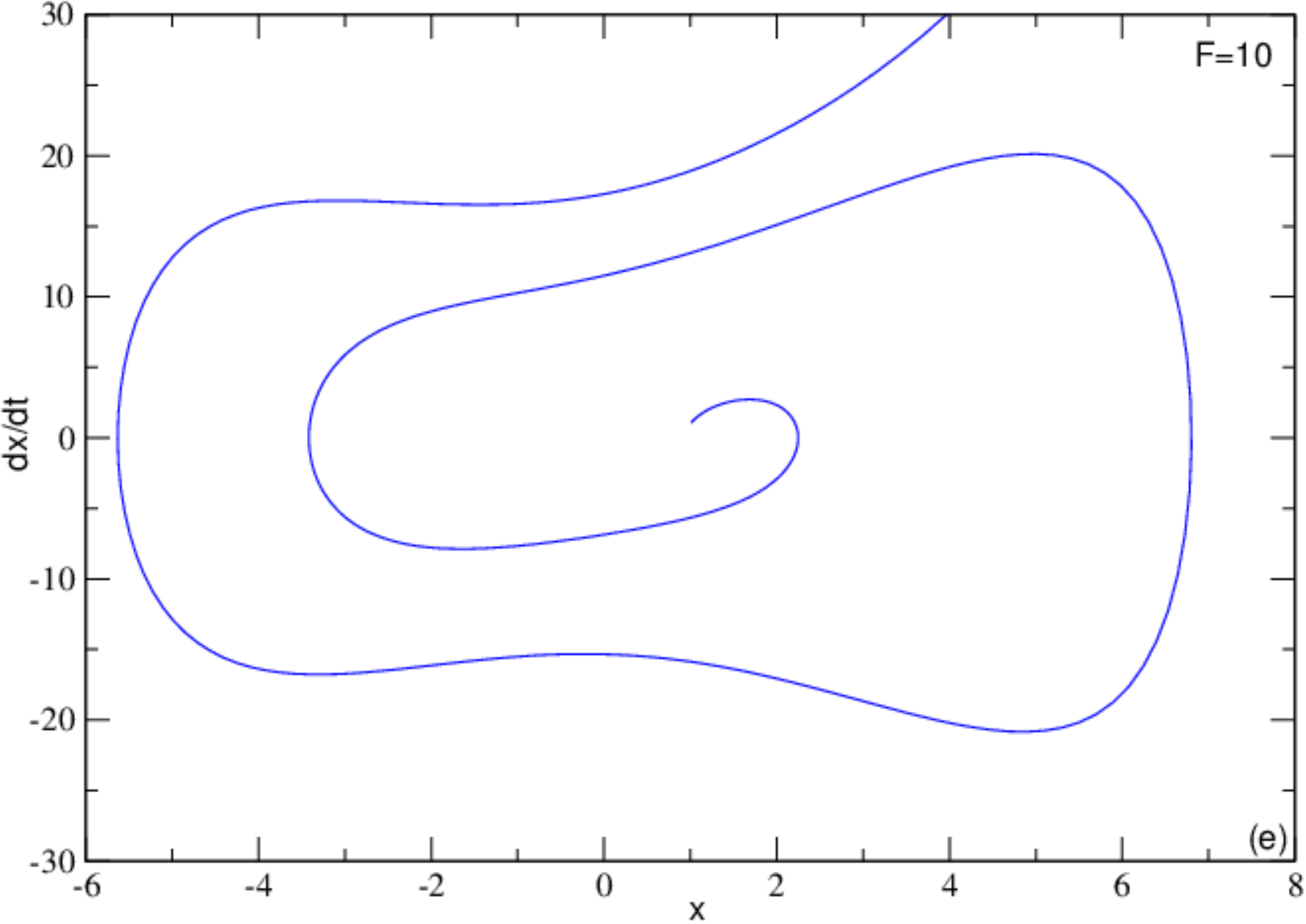}
\end{center}
\caption{Various phase portraits for several different values of F with the
parameters of Fig.\ref{fig:9}.}
\label{fig:12}
\end{figure}

\begin{figure}[htbp]
\begin{center}
 \includegraphics[width=8cm,  height=6cm]{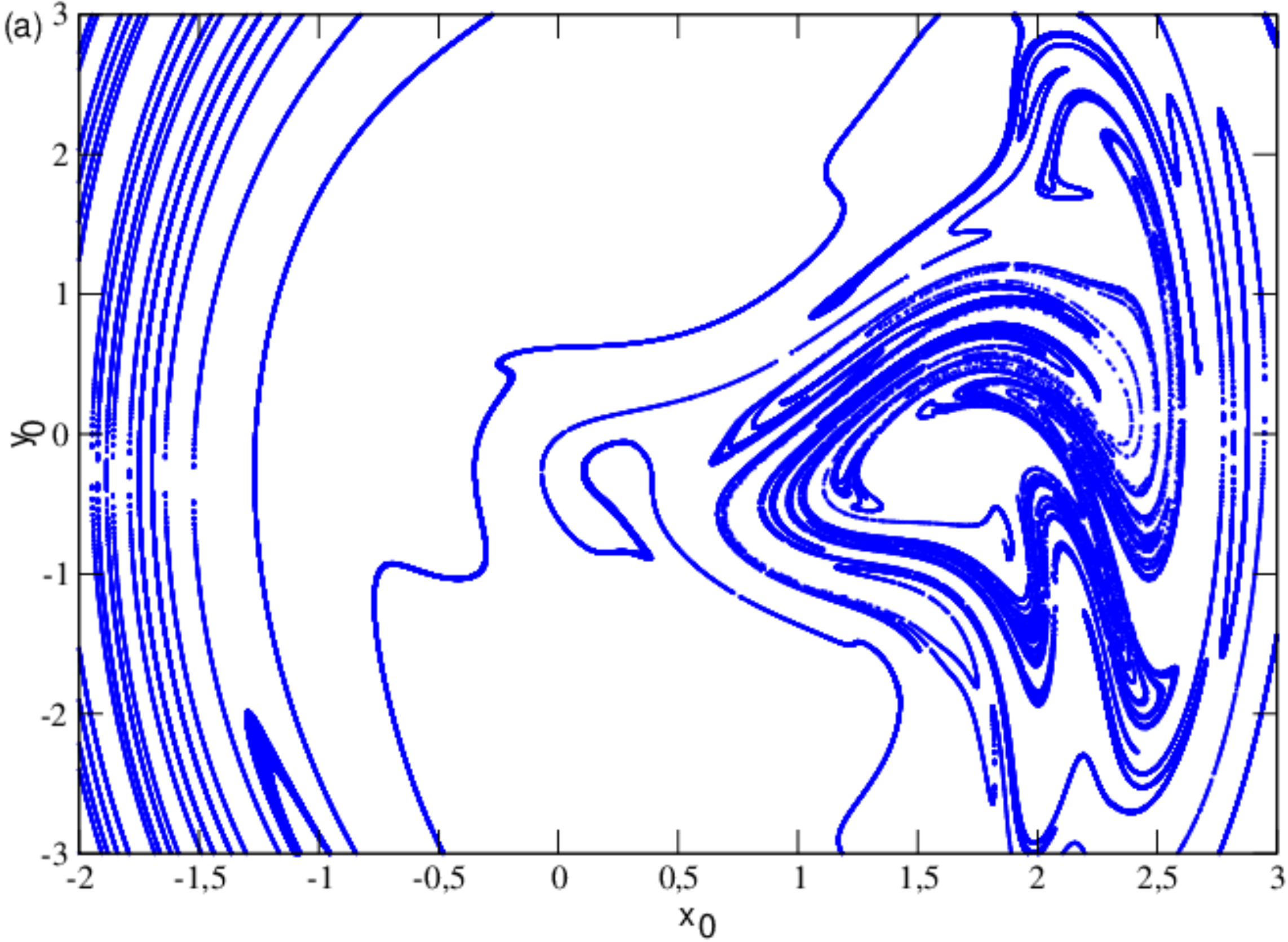}
\includegraphics[width=8cm,  height=6cm]{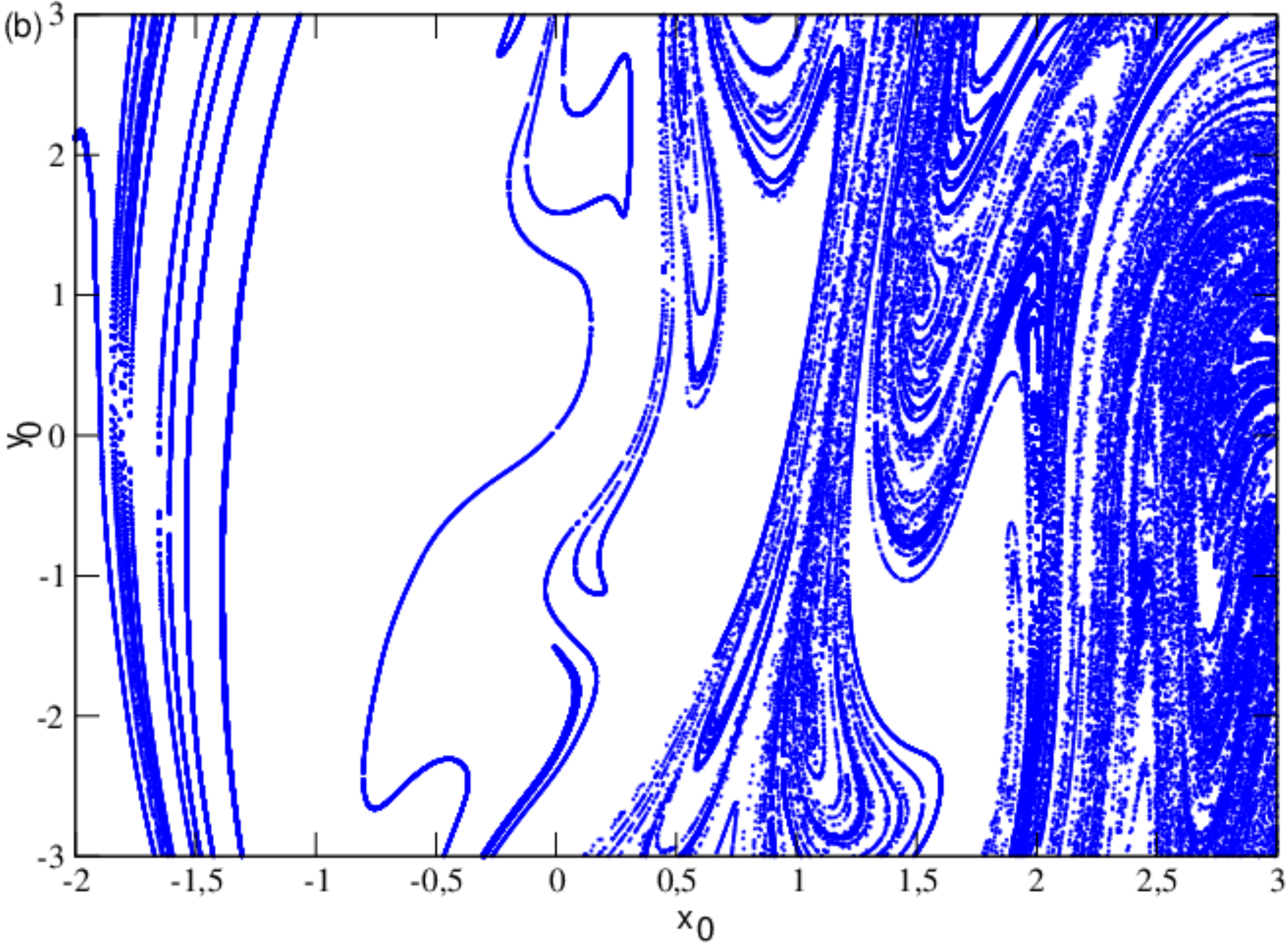}
\includegraphics[width=8cm,  height=6cm]{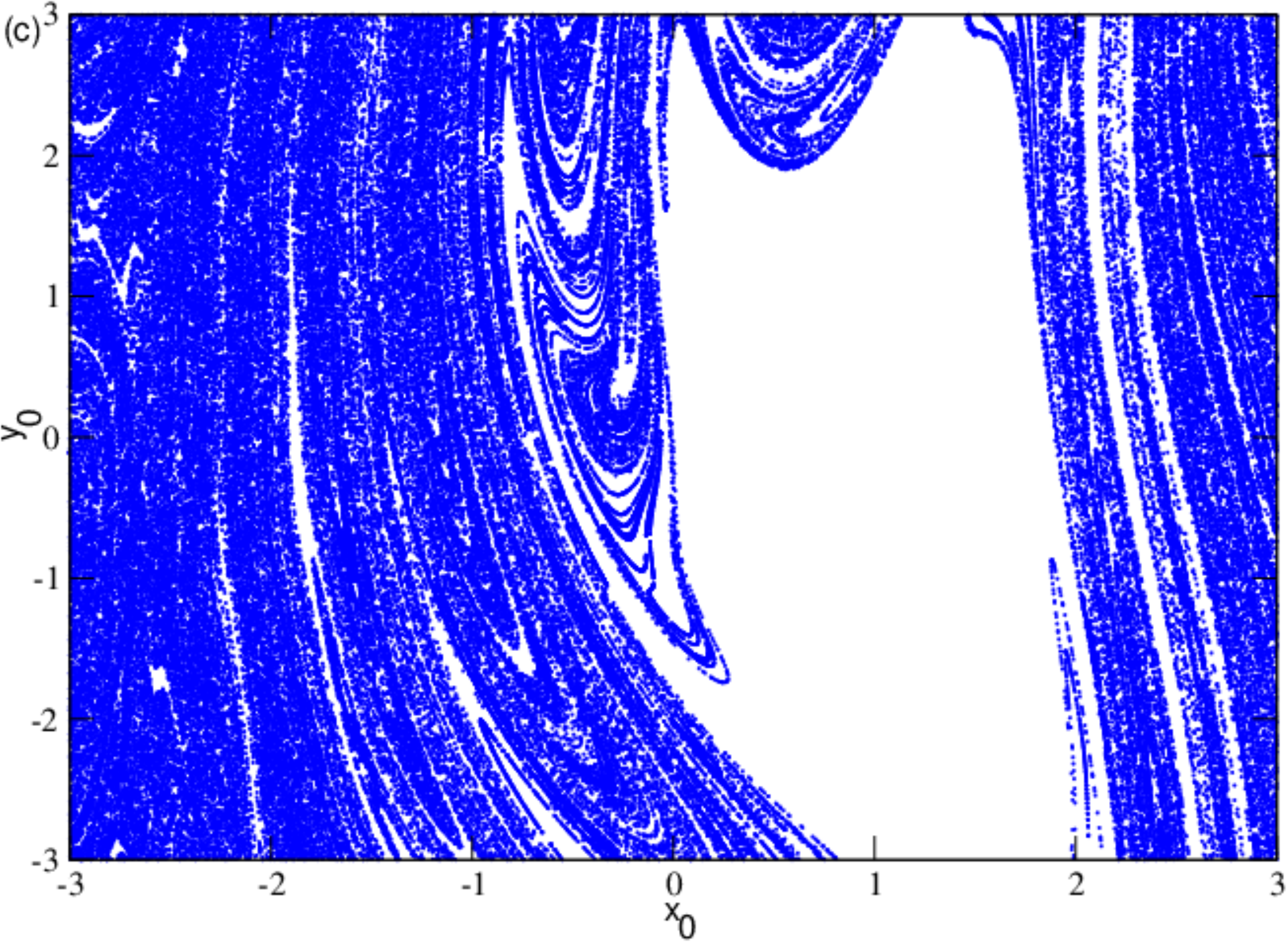}
\end{center}
\caption{Various basin of chaoticity in the primary resonant state with parameter of and $(a) F=5, (b) F=10$ and $(c) F=15$.}
\label{fig:13}
\end{figure}

\begin{figure}[htbp]
\begin{center}
 \includegraphics[width=8cm,  height=8cm]{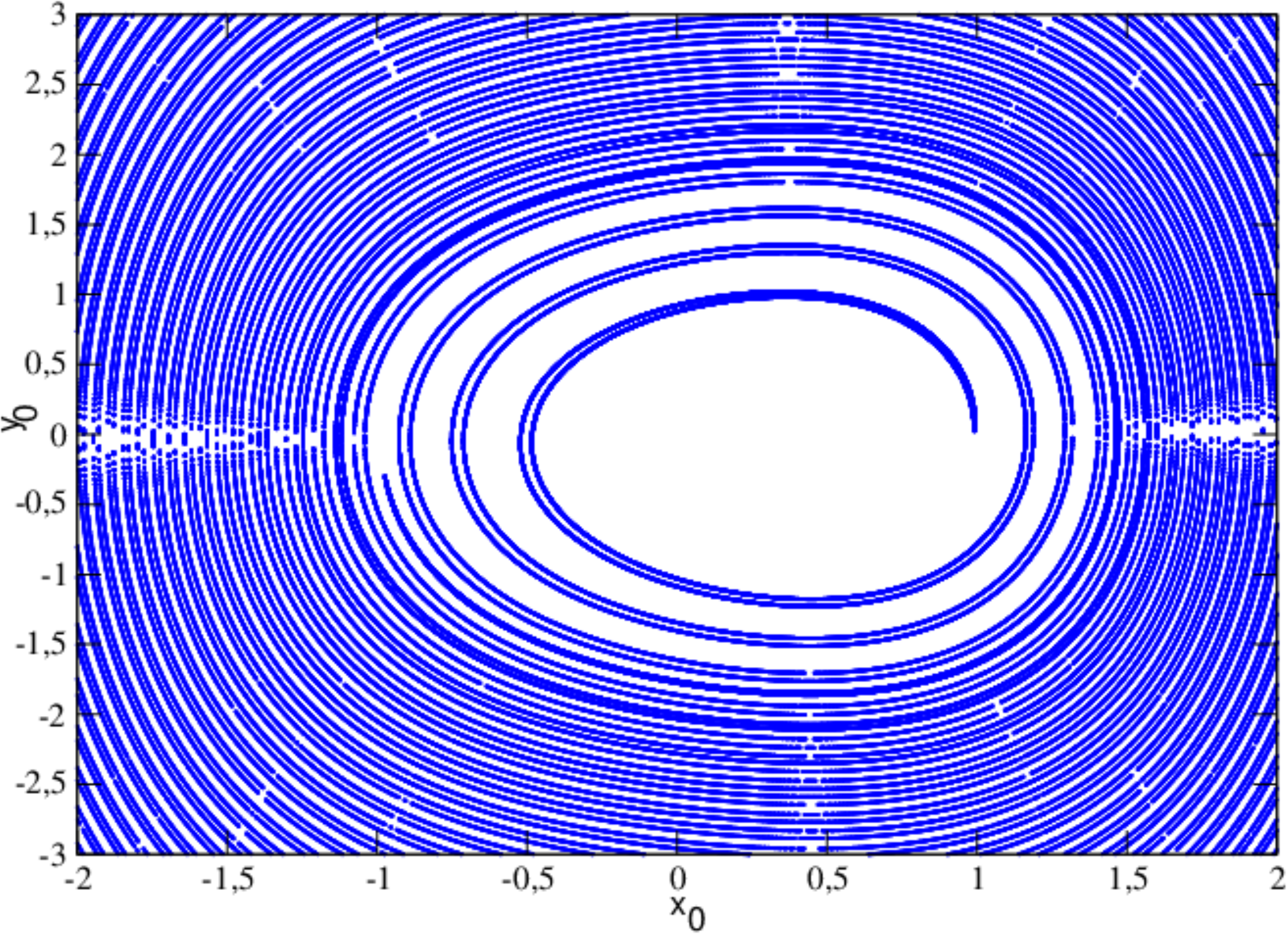}
\end{center}
\caption{Basin of chaoticity in the superharmonic resonant state.}
\label{fig:14}
\end{figure}

\begin{figure}[htbp]
\begin{center}
 \includegraphics[width=8cm,  height=8cm]{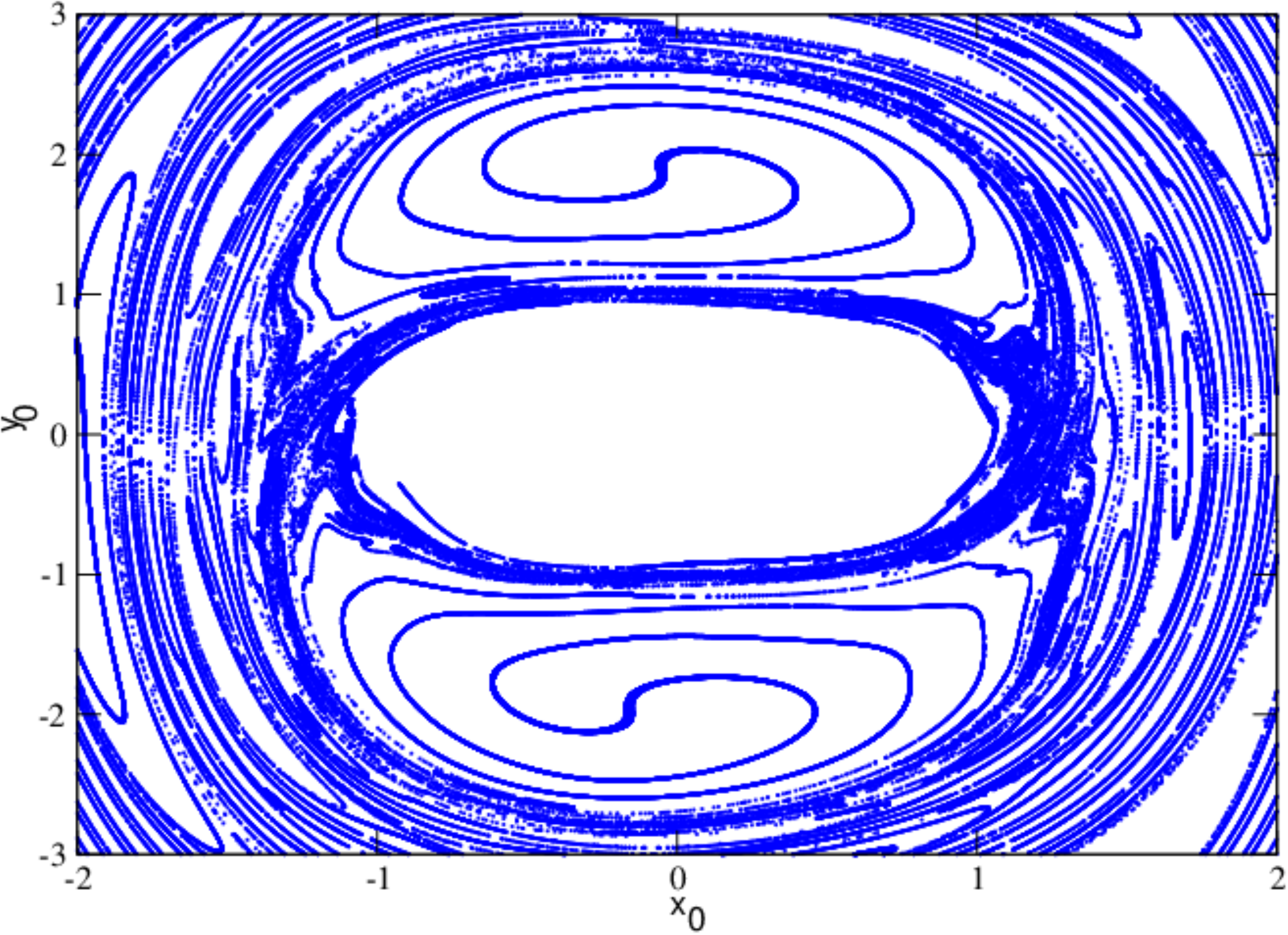}
\end{center}
\caption{Basin of chaoticity in the  subharmonic resonant state.}
\label{fig:15}
\end{figure}

Our aim in this section is to investigate the way under which chaotic motions arise in the model 
described by Eq. (\ref{eq.0}) for resonant states since they are of interest in $(ENSO)$, $MEMS$ device  consisting 
of a $30\mu m$ diameter silicon disk which can be made to vibrate by heating, Brusselator and Selkov oscillators etc. For this purpose,  
we numerically solve this equation using the fourth-order Runge Kutta algorithm \cite{27} and plot the resulting bifurcation
 diagrams and the variation of the corresponding largest Lyapunov exponent as the amplitude $F$,  the parameters of nonlinearity 
$\mu, \beta, k_1, k_2, \lambda$ and $\gamma$ $ \alpha$ are varied. The stroboscopic time period used to map various transitions
which apper in the model is $ T=\frac{2\pi}{\Omega}$. 
The largest Lyapunov exponent which is used here as the instrument to measure the rate of chaos in the system is defined as

\begin{equation}
 Lya=\lim_{t \rightarrow \infty}\frac{ln \sqrt{dx^2+d\dot x^2}}{t}\label{eq70}
\end{equation}
where $dx$ and $d\dot x$ are respectively the variations of $x$ and $\dot x$.
Initial condition that we are used in the simulations of this section is $(x_0, \dot x_0)=(1, 1)$.
In order to have an idea about the system behavior as predicted by the bifurcation diagram, various phase
portraits for several different values of F chosen in the above mentioned regions are plotted in
Figs. \ref{fig:10}, \ref{fig:11} and \ref{fig:12} using respectively  the parameters of Figs. \ref{fig:7}, \ref{fig:8} and \ref{fig:9}.
It should be emphasized from Figures \ref{fig:7}, \ref{fig:8} and \ref{fig:9} that there 
are some domains where the Lyapunov exponent does not match very well the regime of oscillations
 expected from the bifurcation diagram. Far from being an error which has occurred from
the numerical simulation process, such a behavior corresponds to what is called the intermittency
phenomenon. Therefore, within these intermittent domains, the dynamics of the model can not be
predicted. For instance, some forecasted period-1 and quasiperiodic motions from the bifurcation
diagram are not confirmed by the Lyapunov exponent. Indeed,  phase
portraits display rather quasiperiodic, nonperiodic or chaotic motions. Due to
the high sensitivity of the model to initial conditions, basins of chaoticity (where for any choice
of initial conditions that belongs to the shaded area will lead the system to chaotic states while if
the initial conditions are chosen in the non-shaded area, the system will display periodic or quasi-
periodic states) are also checked in primary, superharmonic and subharmonic resonant states (see
Figs. \ref{fig:13}, \ref{fig:14} and \ref{fig:15}. respectively). From these figures, we conclude that chaos is more abundant in
the superharmonic resonant states than in the  primary and subharmonic resonances. This confirms
what has been obtained through their bifurcation diagrams and Lyapunov exponent. To show how the parameters of modified nonlinearity
can influence the chaotic motion in the model, the Lyapunov exponent has also been plotted versus $\mu, k_1,  k_2,\beta$ and parametric 
excitation amplitude with the parameters of Fig. \ref{fig:7} and the following results are observed: As

\newpage
\section{Conclusion}
We have investigated regular and chaotic behaviors of modified Rayleigh-Duffing oscillator. The  amplitude of harmonic amplitude is
found by using the balance harmonic method. It is obtained the hysteresis and jump phenomena and resonance phenomenon have 
appeared in $(\Omega, A)$ space. It is found that the nonlinear damping and parametric excitation amplitude affected severaly the amplitude and frequency of resonance curve. 
Various bifurcation structures showing different types of transitions from quasi-periodic motions to
periodic, unperiodic and chaotic motions have been drawn and the influences of different parameters on these motions have been study. It is noticed 
that chaotic motions have been controlled by the  parameters $\mu,  \beta, k_1, k_2, \alpha$,  and external frequency.
The results of basin attraction show a way to predict initial conditions which regular and chaotic behaviors are obtained. This could be helpful for 
experimentalists who are interested in trying to stabilize such a system with differents parameters or  initial conditions
. For practical interests,  it is useful to develop tools
 and to find ways to control or suppress such undesirable regions. This will be also  useful to control high amplitude of oscillations obtained 
and which are generally source of instability in systems which modeled by this modified Rayleigh-Duffing oscillator equation.

\section*{Acknowlegments}
The authors thank IMSP-UAC and Benin gorvernment for financial support. We also thank Laurent Hinvi for his fruitfull suggestions and
Professor Paul Woafo for his suggestions and collaboration.


\begin{thebibliography}{100}
\bibitem{1}
Nayfeh, A. H., and Mook, 
\emph{D. T., 1979, Nonlinear Oscillations},  Wiley, New York.
\bibitem{2}
Hayashi. C.,
\emph{1964, Nonlinear Oscillations in Physical Systems}, McGraw-Hill, New York.
\bibitem{3}
Chedjou,J. C., Fostin, H. B., and Woafo, P.,
\emph{ 1997, ''Behavior of the van der Pol Oscillator with Two External Periodic forces''},Phys. Scr., 55,pp,390-393. DEA de Physique des liquides. Paris VI- Ecole Polytechnique.
\bibitem{4}
Carrol, T. L.,
\emph{1995, ''Communicating With Use of filtered, Synchronized, Chaotic Signals.''} IEEE Trans, Circuits syst.,I: Fundam. Theory Appl.,42,pp. 105-110. 

\bibitem{5}  
Alberto Francescutto, Giorgio Contento,
\emph{ Bifurcations in ship rolling: experimental results
and parameter identification technique},  Ocean Engineering 26 (1999) 1095–1123.
\bibitem{6} 
K.W. Holappa, J.M. Falzarano
\emph{Application of extended state space to nonlinear
ship rolling}, Ocean Engineering 26 (1999) 227–240.

\bibitem{7} 
Wan Wu, Leigh McCue,
\emph{Application of the extended Melnikov’s method for single-degree-of-freedom
vessel roll motion}, Ocean Engineering 35 (2008) 1739–1746.

\bibitem{8} R. A. Mahaffey,  
\emph{Physics of Fluids} $19$,  $1837$ ($1976$).

\bibitem{10} K. Ostrikov and S. Xu,  
\emph{Plasma-aided Nanofabrication: from Plasma Sources to 
Nanoassem-bly} (John Wiley Sons,  Weinheim,  $2007$),  pp. $149-280$.
\bibitem{11} H. G. Enjieu Kadji,  J. B. Chabi Orou and P. Woafo,  
\emph{Physica Scripta} (In Press) ($2007$).

\bibitem{12}H. G. Enjieu Kadji,  B. R. Nana Nbendjo,  J. B. Chabi Orou,  and P. K. Talla, 
\emph{ Nonlinear dynamics of plasma oscillations modeled by an anharmonic oscillator} ($2007$).
\bibitem{13}
S.H.Strogatz,  
\emph{Nonlinear dynamics and chaos with applications to physics,  chemistry and engineering},  (Westview Press,  Cambridge,  $1994$),  Sec. $1.2$.
\bibitem{14} 
R. Yamapi, M.A. Aziz-Alaoui,
\emph{Vibration analysis and bifurcations 
in the self-sustained electromechanical system with multiple functions},
Communications in Nonlinear Science and Numerical Simulation $12 (2007) 1534-1549$.

\bibitem{15}
Darya V. Verveyko and Andrey Yu. Verisokin,
\emph{Application of He's method to the modified Rayleigh equation},Discrete and Continuous Dynamical Systems, Supplement 2011, pp. 1423–1431.
\bibitem{16}
Aubin, K., Zalalutdinov, M., Alan, T., Reichenbach, R. B., Rand, R. H., Zehnder, A.,
Parpia, J. and Craighead, H. G.,
\emph{Limit Cycle Oscillations in CW Laser-Driven NEMS,}
Journal of Micro-electrical mechanical System 13:1018-1026, 2004.
\bibitem{17}
Zalalutdinov, M., Olkhovets, A., Zehnder, A., Ilic, B., Czaplewski, D. and Craighead, H.
G.,
\emph{Optically pumped parametric amplification for micro-mechanical systems,} Applied
Physics Letters 78:3142-3144, 2001.
\bibitem{18}
Rand, R. H., Ramani, D. V, Keith, W. L. and Cipolla, K. M.,
\emph{The quadratically damped
Mathieu equation and its application to submarine dynamics,} Control of Vibration and
Noise: New Millennium 61:39-50, 2000.
\bibitem{19}
Wirkus, S., Rand, R. H. and Ruina, A., 
\emph{How to pump a swing,} The College Mathematics
Journal 29:266-275, 1998.
\bibitem{20}
Zhehe Y.,Deqing M., Zichen C., 
\emph{Chatter suppression by parametric excitation: Model
and experiments,} Communications in Nonlinear Science and Numerical Simulation
330:2995-3005, 2011.

\bibitem{21}
Wang, B. and Fang, Z.,
 
\emph{‘Chaotic Oscillations of Tropical Climate: A Dynamic
System Theory for ENSO’,
} Journal of Atmospheric Sciences 53:2786-2802, 1996.


\bibitem{22}
 Wang, B., Barcilon, A. and Fang, Z.,

\emph{‘Stochastic Dynamics of El Nino-
Southern Oscillation’, 
}  Journal of Atmospheric Sciences 56:5-23, 1999.

\bibitem{23}
 Zalalutdinov, M., Parpia, J.M., Aubin, K.L., Craighead, H.G., T.Alan, Zehn-
der, A.T. and Rand, R.H., 
\emph{Hopf Bifurcation in a Disk-Shaped NEMS, Pro-
ceedings of the 2003 ASME Design Engineering Technical Conferences,
} 19th
Biennial Conference on Mechanical Vibrations and Noise, Chicago, IL, Sept.
2-6, paper no.DETC2003-48516, 2003 (CD-ROM).

\bibitem{24}
 Pandey, M., Rand, R. and Zehnder, A.,
\emph{‘Perturbation Analysis of Entrainment
in a Micromechanical Limit Cycle Oscillator’
},Communications in Nonlinear
Science and Numerical Simulation, available online, 2006.

\bibitem{25}
Tina Marie Morrison, 
\emph{Three problems in nonlinear dynamics with 2:1 parametric excitation},  Ph.D. Cornell University 2006.

\bibitem{26} A. H. Nayfeh,  
\emph{ Introduction to perturbation techniques},  
(John Wiley and Sons,  New York,  $1981$), 
Sec.$4.5$.
\bibitem{27} N. Piskunov,  
\emph{Calcul Différentiel et intégral},  Tome II,  $9^e$ edition,  MIR,  Moscou (1980).
\bibitem{28}
Soliman, M. S. and Thompson, J. M. T. 
\emph{The effect
of damping on the steady state and basin bifurcation
patterns of a nonlinear mechanical oscillator,} Int. J.
Bifurcation and Chaos 2, 81–91 (1992).
\bibitem{29} 
 Sanjuan MAF. 
\emph{ The effect of nonlinear damping on the universal escape oscillator}. Int J Bifurcat Chaos 1999;9:735.


\end{thebibliography}
\end{document}